\documentclass[twocolumn,superscriptaddress,longbibliography,prc, nofootinbib]{revtex4-2}
\usepackage{times}
\usepackage[T1]{fontenc}

\usepackage{physics}
\usepackage{amsmath,amssymb,mathrsfs}
\usepackage{graphicx}
\usepackage{dcolumn}
\usepackage{bm}
\usepackage{morefloats}
\usepackage{multirow}
\usepackage{amssymb}
\usepackage{amsmath}
\usepackage{xcolor}
\usepackage{longtable}
\usepackage{fix-cm}
\usepackage{verbatim}
\usepackage{float}
\usepackage{mathtools}
\usepackage{mathptmx} 
\usepackage[T1]{fontenc}
\usepackage[colorlinks,allcolors=blue]{hyperref}

\makeatletter
\def\NAT@def@citea{\def\@citea{\NAT@separator}}
\makeatother

\def\0\\{\nonumber\\}

\newcommand{\beq}{\begin{equation}}
\newcommand{\eeq}{\end{equation}}
\newcommand{\beqn}{\begin{eqnarray}}
\newcommand{\eeqn}{\end{eqnarray}}

\newcommand{\kfx}{k_{\mathrm{Fx}}}

\newcommand{\Deltamax}{\Delta_{\mathrm{max}}}
\newcommand{\kmax}{k_{\mathrm{max}}}

\newcommand{\nb}{n_{\mathrm{B}}}

\makeatletter
\newcommand\footnoteref[1]{\protected@xdef\@thefnmark{\ref{#1}}\@footnotemark}
\makeatother

\begin{document}


\title{Data-driven exploration of the neutron $^3\text{P}_2$ pairing gap using Cassiopeia A neutron star observational data: Direct $\chi^2$ minimization}

\author{Yoonhak Nam}
\email{nam.y.b76c@m.isct.ac.jp}
\affiliation{Department of Physics, School of Science, Institute of Science Tokyo, Tokyo 152-8551, Japan}

\author{Kazuyuki Sekizawa}
\email{sekizawa@phys.sci.isct.ac.jp}
\affiliation{Department of Physics, School of Science, Institute of Science Tokyo, Tokyo 152-8551, Japan}
\affiliation{Nuclear Physics Division, Center for Computational Sciences, University of Tsukuba, Ibaraki 305-8577, Japan}
\affiliation{RIKEN Nishina Center, Saitama 351-0198, Japan}

\date{\today}

\begin{abstract}
\edef\oldrightskip{\the\rightskip}
\begin{description}
\rightskip\oldrightskip\relax
\setlength{\parskip}{0pt} 

\item[Background]
The rapid cooling observed in the Cassiopeia~A neutron star (Cas~A NS) provides one of the most stringent tests for neutron star cooling theory. While the Cooper-pair breaking and formation (PBF) neutrino emission process is a leading candidate, significant theoretical uncertainties remain regarding both the PBF efficiency factor $q$ and the neutron $^{3}\mathrm{P}_{2}$ pairing gap.

\item[Purpose]
This work aims to explore, in a data-driven manner, how the optimized shape of the neutron ${}^{3}\mathrm{P}_{2}$ pairing gap responds to variations of the phenomenological PBF emissivity parameter~$q$ within a fixed and controlled cooling setup.

\item[Methods]
We introduce a novel parametrization of the neutron ${}^{3}\mathrm{P}_{2}$ pairing gap, in which each parameter carries direct physical meaning, allowing independent control of the gap amplitude, peak location, width, and asymmetry. Using a newly developed Fortran-based neutron star cooling code and the BSk24 equation of state, we perform systematic parameter-space exploration guided by the Cas~A NS observational data. Global optimization is carried out with Optuna’s tree-structured Parzen estimator (TPE), followed by local refinement using the Nelder-Mead simplex method. Both single-objective optimizations based solely on $\chi^{2}$ and multi-objective optimizations that combine $\chi^{2}$ with a slope-difference metric are examined under identical conditions.

\item[Results]
The optimized solutions yield physically reasonable neutron ${}^{3}\mathrm{P}_{2}$ pairing gaps with peak amplitudes $\Delta_{\max}\approx0.5$--$0.6~\mathrm{MeV}$. While the multi-objective formulation explores the parameter space more broadly,
the single-objective ($\chi^{2}$-only) optimization finally achieves the lowest $\chi^{2}$, reflecting the intrinsically curved nature of neutron star cooling trajectories and the fact that minimizing a slope difference at a single epoch does not necessarily guarantee the best global agreement with the full Cas~A NS dataset. Adopting the canonical neutron star mass $M_{\mathrm{NS}}=1.4\,M_\odot$, we find that increasing $q$ drives the optimized gap and critical-temperature profiles toward smoother and more localized shapes and improves consistency with the observed cooling trend. In particular, models with $q\gtrsim0.4$ reproduce the Cas~A NS decline rate within the $1\sigma$ confidence interval, whereas smaller values of $q$ lead to progressively shallower cooling slopes, with the baseline case $q\simeq0.19$ lying near the $3\sigma$ level within the present setup.

\item[Conclusions]
Within this restricted and controlled modeling framework, our results suggest larger effective PBF emissivities than the baseline estimate. However, a statistically robust and less assumption-dependent constraint on $q$ requires a dedicated Bayesian inference that simultaneously accounts for uncertainties in neutron star mass, envelope composition, equation of state, pairing microphysics, and the age offset of Cas~A~NS. The physically interpretable gap parametrization and hybrid optimization strategy developed here provide a natural foundation for such future Bayesian and machine-learning-based studies of neutron star cooling.

\end{description}
\end{abstract}

\maketitle

\section{Introduction}\label{Sec:Intro}

The neutron star within the supernova remnant Cassiopeia A (Cas~A NS), discovered by the Chandra X-ray Observatory in 1999 \cite{Hughes_2000}, represents one of the most intensively studied compact objects in modern astrophysics. Based on kinematic analysis of the supernova remnant, this neutron star was formed through a supernova explosion that occurred in 1681$\pm$19 \cite{Fesen_2006}, making it approximately 340 years old and thus one of the youngest known neutron stars. Unlike other known neutron stars that exhibit complex magnetospheric activity, Cas~A NS belongs to the X-ray thermal isolated neutron stars (XTINSs), emitting purely thermal soft X-ray radiation without detectable radio or gamma-ray emission \cite{Pavlov_2009_no_pulsation}. This thermal simplicity, combined with its young age, provides an exceptionally clean laboratory for studying neutron star cooling physics under well-constrained conditions.

Among known neutron stars, Cas~A NS holds a unique position as the only isolated compact object for which long-term thermal monitoring has been continuously conducted across multiple decades. This exceptional observational baseline provides an unprecedented window into real-time stellar evolution processes during the critical early phases of neutron star thermal development. The continuous monitoring has revealed a measurable decline in surface temperature that can be observed in real-time, making Cas~A NS an invaluable test case for theoretical cooling models.

Over more than two decades of monitoring, multiple research groups have employed different observational strategies and analysis techniques to characterize the thermal evolution of Cas~A NS. Early studies utilized Chandra's Advanced CCD Imaging Spectrometer (ACIS) in Graded mode, primarily designed for supernova remnant observations \cite{Heinke_2010_direct_observation_1, Elshamouty_2013_direct_observation_2}. Subsequently, dedicated observations using the Faint mode were advocated to minimize instrumental effects such as photon pileup \cite{Posselt_2018_upper_limit_1,Posselt_2022_upper_limit_2}. Recent comprehensive analyses have attempted to reconcile data from both observational modes through careful calibration procedures \cite{Shternin_2022_1.55_0.25}, although systematic uncertainties persist in the derived cooling parameters and continue to challenge precise theoretical interpretations.

The observed rapid cooling rate of Cas~A NS significantly exceeds predictions from conventional neutron star cooling scenarios dominated by modified-Urca neutrino emission processes, highlighting the rapid cooling of the Cas~A NS. This fundamental discrepancy has prompted the scientific community to explore a diverse range of alternative mechanisms to explain the enhanced cooling behavior. Proposed explanations include thermal recovery following r-mode activity \cite{Yang_2011_r_mode}, rotation-driven particle repopulation that triggers direct-Urca cooling \cite{Negreiros_2013_rotation_durca}, strong suppression of thermal conductivity by medium effects \cite{Blaschke_2012_thermal_cond_medium,Blaschke_2013_thermal_cond_medium, Taranto_2016_medium_durca}, magnetic-field-decay (Joule) heating \cite{Bonanno_2014_joule}, superfluid quantum criticality effects \cite{zhu_2024_quantum_criticality}, exotic particle emission such as axions \cite{Leinson_2014_axion,Leinson_2021_axion, Sedrakian_2016_axion,Sedrakian_2019_axion, Hamaguchi_2018_axion}, and phase transitions in putative quark-matter cores \cite{Noda_2013_quark,Sedrakian_2013_quark, Sedrakian_2016_quark} (for more details, see a recent review~\cite{potekhin_2025_durca_in_cas_a} and references therein).

Among these diverse theoretical proposals, the Cooper-pair breaking and formation (PBF) mechanism---also referred to as Cooper-pair formation (CPF) in the neutron star cooling literature---has emerged as one of the most physically motivated explanations; for brevity, we use PBF throughout this article. The PBF process occurs when neutrons form Cooper pairs and transition to a superfluid state as the neutron star core temperature drops below the critical temperature. During this transition, pre-existing Cooper pairs are broken and reformed under thermal fluctuations, mediated through weak neutral currents that emit neutrino pairs. This mechanism is naturally activated when superfluidity onset occurs as the core temperature gradually decreases with increasing neutron star age, providing a natural explanation for the timing of the observed cooling acceleration.

The theoretical foundation for PBF cooling was initially established in the 1970s and 1980s \cite{Flowers_1976_PBF, voskresensky_1987_PBF}, with subsequent refinements by multiple research groups over several decades. The neutrino emissivity from the PBF process is expressed in the following general form \cite{Yakovlev_2001_neutrino_emissivity, Schmitt_2018_neutrino_emissivity}:
\begin{equation}
Q_{\text{PBF}} = q \cdot Q_{\text{PBF0}} \cdot T^7 \cdot \mathcal{F}(v), \label{emissivity_of_PBF}
\end{equation}
where $Q_{\text{PBF0}}$ is a temperature-independent prefactor determined by the fundamental material properties and neutrino interaction constants, expressed as:
\begin{equation}
Q_{\text{PBF0}} = 1.17 \times 10^{-42} \left(\frac{m_\text{n}^*}{m_\text{N}}\right) \left(\frac{p_{\mathrm{Fn}}}{m_\text{N} c}\right) N_\nu a_n  \,\,\mathrm{erg\,cm^{-3}\,s^{-1}\,K^{-7}}.
\end{equation}
In this expression, $N_\nu = 3$ is the number of neutrino flavors, $m_\text{n}^*$ is the neutron effective mass at the Fermi surface, $p_{\mathrm{Fn}}$ is the neutron Fermi momentum, and $m_\text{N}$ is the bare nucleon mass. The numerical constant $a_n = g_V^2 + 2g_A^2 \simeq 4.17$ encompasses contributions from the vector coupling constant ($g_V \simeq 1$) and axial-vector coupling constant ($g_A \simeq 1.26$) of the weak interaction. The auxiliary function $\mathcal{F}(v)$ depends on the dimensionless gap parameter $v = \Delta_0/(k_B T)$, where $\Delta_0$ represents the neutron triplet gap amplitude. An analytical approximation for this function can be found in Ref.~\cite{Yakovlev_2001_neutrino_emissivity}. The phenomenological efficiency factor $q$ in Eq.~(\ref{emissivity_of_PBF}) accounts for many-body corrections, the most prominent of which relates to the response of the superfluid condensate.

Early formulations suggested substantial neutrino emissivity from both vector and axial current channels in superfluid matter. However, critical theoretical advances revealed that vector current contributions suffer from relativistic suppression factors due to the requirement of vector current conservation, effectively eliminating singlet pairing contributions \cite{Leinson_2006_vector_current}.

In Ref.~\cite{Page_2009_vector_channel}, Page \textit{et al.} proposed a phenomenological correction suggesting that neutron $^3\text{P}_2$ superfluidity completely suppresses the vector channel, setting the phenomenological efficiency factor in Eq.~\eqref{emissivity_of_PBF} as:
\begin{equation}
q = \frac{2g_A^2}{a_n} = \frac{2g_A^2}{g_V^2 + 2g_A^2} \simeq 0.76.
\end{equation}
This correction was utilized in several Cas~A NS cooling scenario studies \cite{Page_2016_NSCool,shternin_2011_nt_SYHHP, Wijngaarden_2019_diffusive_burning}. 
However, the most recent and comprehensive theoretical treatment \cite{Leinson_2010} has introduced significant complications for the PBF cooling scenario. Advanced microscopic calculations considering the response effects of order parameters in the axial-vector channel revealed that an additional suppression factor of 4 occurs even in the triplet case. This yields the following efficiency factor in the non-relativistic limit:
\begin{equation}
q = \frac{g_A^2}{2g_V^2 + 4g_A^2} \simeq 0.19.
\end{equation}
The resulting theoretical efficiency factor of approximately 0.19, compared to the previously used value of 0.76, appears insufficient to reproduce the observed Cas~A NS cooling rate according to detailed stellar evolution simulations \cite{shternin_2011_nt_SYHHP, Ho_2021_q_0.19_insufficient, Shternin_2022_1.55_0.25}\footnote{
We will often express to these $q$ values simply as 0.76 and 0.19, respectively, although precise values involve smaller digits.
}.

We should note, as pointed out in Ref.~\cite{shternin_2011_nt_SYHHP}, that Leinson's calculations \cite{Leinson_2010} were performed in the non-relativistic limit, and the effects of relativistic corrections on the results remain unclear. Additionally, there may be further modifications due to condensate reaction effects and other possible corrections from collective many-body correlations. Recognizing these theoretical limitations, numerous studies have adopted an approach treating $q$ as an observationally determined free parameter (\textit{e.g.}, Refs.~\cite{Ho_2021_q_0.19_insufficient, Shternin_2022_1.55_0.25, zhu_2024_quantum_criticality}). 

Indeed, recent observational analyses have shown that the efficiency factor of the PBF process must be $q \gtrsim 0.4$ at 90\% confidence level \cite{Ho_2021_q_0.19_insufficient}, and in the range $q = 0.5$--$2.6$ (for the variable effective hydrogen column density $N_\mathrm{H}$ case) and $q = 0.4$--$2.1$ (for the fixed effective hydrogen column density $N_\mathrm{H}$ case) at 68\% confidence level \cite{Shternin_2022_1.55_0.25}. These values are at least 2--3 times higher than theoretical prediction in Ref.~\cite{Leinson_2010} of $q \simeq 0.19$, revealing a serious discrepancy between current microscopic calculations and observational data. Additionally, the maximum critical temperature of neutron $^3\text{P}_2$ pairing has been constrained to the range $T_\text{Cn}^\text{max} = (4-9.5) \times 10^8$ K \cite{Shternin_2022_1.55_0.25}, and these results have been demonstrated to be robust across various equations of state and superfluidity models.

To address this discrepancy, two alternatives that do not rely on enhanced PBF emissivity have been advanced. First, a hybrid cooling picture posits that Cas~A NS's mass lies just above the direct-Urca threshold, so a tiny central direct-Urca kernel has been present since birth while PBF operates in the surrounding core; the joint action of PBF plus a small direct-Urca core reproduces the observed decline without artificially boosting PBF \cite{Leinson_2022_hybrid_cooling}. Second, a Urca-only interpretation---dispensing with PBF as a dominant channel---argues that the combined action of direct-Urca and modified-Urca, including the in-medium enhancement of the modified-Urca rates near the direct-Urca threshold \cite{Shternin_2018_enhanced_murca}, can match the Cas~A NS cooling trend even with low PBF efficiency ($q\!\simeq\!0.19$), weak proton superfluidity, and a carbon envelope, provided the stellar mass slightly exceeds the direct-Urca threshold so that a small, long-lived cold kernel forms \cite{potekhin_2025_durca_in_cas_a}.

Still, significant theoretical uncertainties remain regarding both the PBF process efficiency and the neutron $^3\mathrm{P}_2$ superfluid gap models themselves. Beyond the inherent uncertainties in theoretical calculations of neutrino emissivity, the density-dependent critical temperatures for superfluidity onset remain poorly constrained by nuclear theory. These combined uncertainties have motivated several studies to introduce scaling factors as free parameters to bridge the gap between theory and observations, though such approaches highlight the need for more systematic theoretical treatments.

In the present work, we adopt a systematic, data-driven optimization framework to explore the interplay between the neutrino PBF emissivity scaling factor $q$ and the neutron ${}^3\mathrm{P}_2$ superfluid gap function $\Delta_{\rm n}(k_{\rm Fn})$ within a fixed cooling setup. Rather than taking a specific gap shape as given, we treat the gap function as flexible (but physically motivated) and optimize its parameters against the Cas~A~NS observations for representative choices of $q$.

As a baseline, we start from the theoretically well-motivated PBF formulation of Leinson \cite{Leinson_2010} (corresponding to $q=0.19$) and examine how much freedom in the
shape of the neutron pairing gap $\Delta_{\rm n}(k_{\rm Fn})$ can be accommodated by the Cas~A~NS data.
We then extend the same optimization procedure to a broader range of $q$ values and map, in a data-driven manner, how the optimized gap shape evolves with $q$ under the same modeling assumptions. Accordingly, the present study is not intended to provide statistically robust constraints on $q$ or on the gap parameters, but rather to offer an exploratory $q$--gap-shape mapping that serves as a preparatory step toward a subsequent Bayesian inference study.

Furthermore, recognizing that existing parameterized gap functions are not well-suited for automated optimization procedures, we introduce a novel parametrization of the superfluid gap function specifically designed for systematic parameter space exploration. This new functional form provides the flexibility needed for robust optimization while maintaining physical consistency with theoretical expectations from nuclear many-body calculations. Furthermore, this parametrization is designed to be suitable for future implementation of machine learning techniques, providing scalability to efficiently handle large-scale parameter space searches and complex nonlinear optimization problems.

This article is organized as follows.
Section~\ref{Sec:Methods} introduces the microphysical inputs and stellar models used in our cooling calculations (BSk24 EoS and TOV structure, envelope treatment, and the adopted
singlet gaps), presents the new energy gap parametrization and its constraints (Sec.~\ref{sf_gap_model}), and describes the data-driven optimization framework, including the Cas~A~NS dataset, radius rescaling, and implementation details.
Section~\ref{Sec:Results} reports our main findings: a comparison between single- and multi-objective formulations, and the $q$-dependence of the optimized gap shape at a fixed canonical neutron star mass of $M = 1.4\,M_\odot$, where $M_\odot$ denotes the mass of the Sun, with accompanying gap and $T_{\mathrm{c}}$ profiles and cooling curve comparisons to the Cas~A~NS data.
Finally, Section~\ref{Sec:Conclusion} summarizes the implications of our exploratory analysis for the PBF efficiency and neutron ${}^{3}\mathrm{P}_{2}$ pairing, outlines possible limitations of the present fixed setup, and discusses extensions toward joint optimization of singlet channels and to Bayesian- and machine-learning-based inference of microphysical and EoS parameters.

\section{Methods}\label{Sec:Methods}

\subsection{Superfluid and superconducting gap models}\label{sf_gap_model}

\begin{figure}[t]
    \centering
    \includegraphics[width=1.0\linewidth]{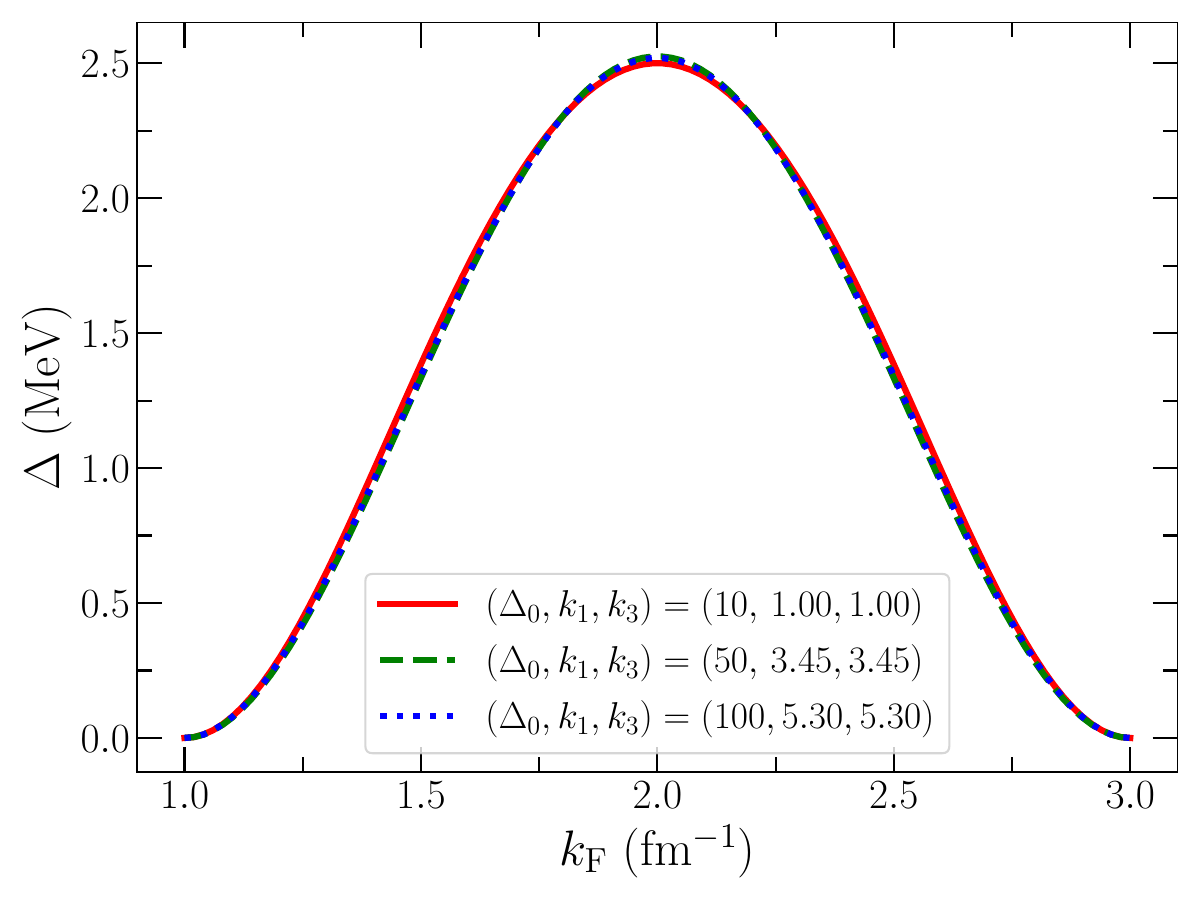}\vspace{-3mm}
    \caption{A figure that shows the degeneracy problem of the conventional pairing gap function (\ref{traditional_parametrization}). Three pairing gap functions with distinct parameter sets are shown as functions of the Fermi wave number. Two parameters, $k_0=1$ and $k_2=3$, are fixed, which are the left and the right edges of the gap functions, respectively. Red solid, green dashed, blue dotted lines correspond to the cases with $(\Delta_0,k_1,k_3)=(10,1,1)$, $(50,3.45,3.45)$, and $(100,5.3,5.3)$, respectively.}
    \label{similar_gaps}
\end{figure}

In neutron star cooling, the PBF processes come into play when neutron star matter cools down below the critical temperatures for superfluidity (neutron $^1\text{S}_0$ or $^3\text{P}_2$) or superconductivity (proton $^1\text{S}_0$). Those critical temperatures has been expressed using the pairing energy gap $\Delta$ as follows:
\begin{align}
k_\mathrm{B}T_{\mathrm{c}}\approx
\begin{dcases}
        0.5669\Delta & \text{singlet~(isotropic pairing) gap}\\
        0.5669\dfrac{\Delta}{\sqrt{8\pi}} & \text{triplet (anisotropic pairing) gap}
\end{dcases}
\label{pairing_gap_temp_relation}
\end{align}
Here, $k_\mathrm{B}$ is the Boltzmann constant. This standard BCS-based convention has been widely used in neutron star cooling studies (see, \textit{e.g.}, Ref.~\cite{Ho_2015}).

\begin{figure*}[t]
    \centering
    \includegraphics[width=1.0\textwidth]{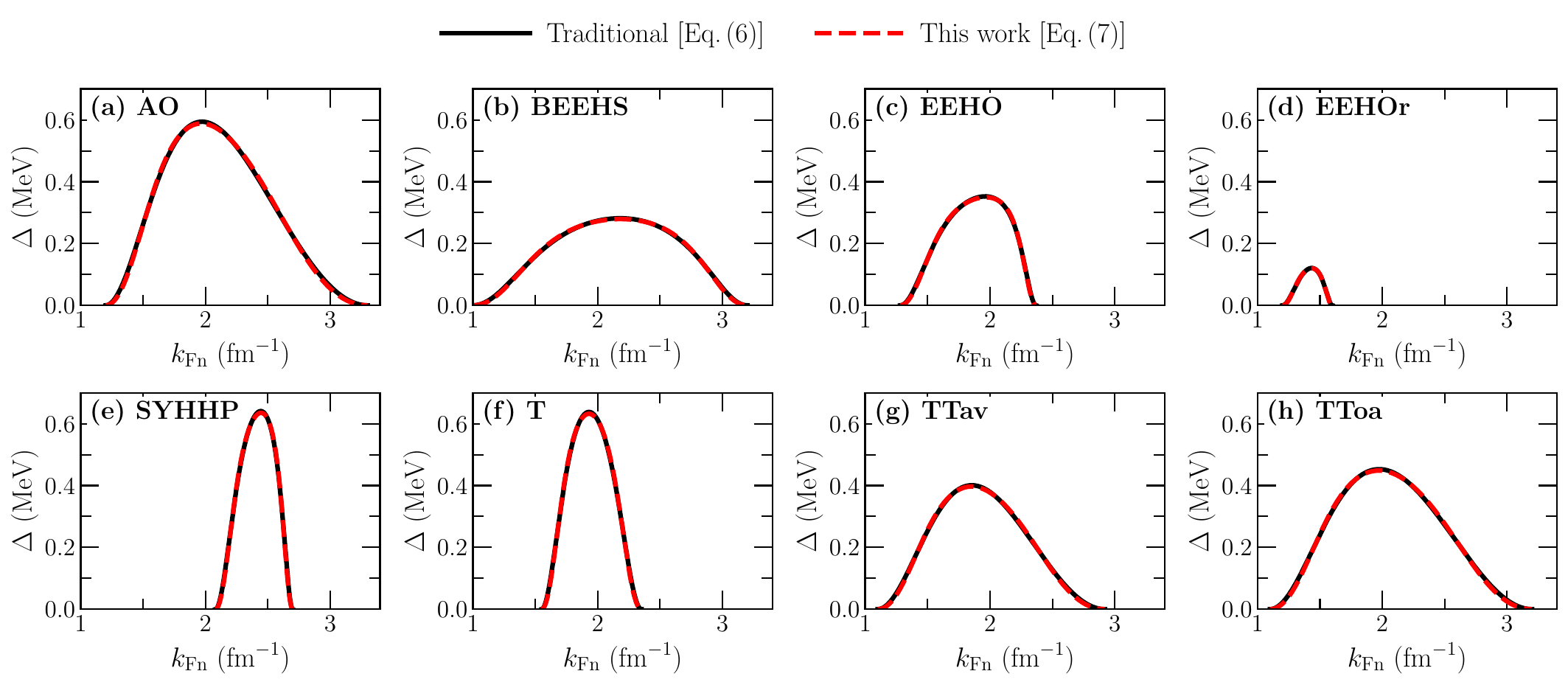}\vspace{-3mm}
    \caption{Comparison of eight commonly used neutron $^3\mathrm{P}_2$ pairing gap models, as listed in Table~II of Ref.~\cite{Ho_2015}---AO, BEEHS, EEHO, EEHOr, SYHHP, T, TTav, and TToa---drawn with the traditional parametrization \eqref{traditional_parametrization} (solid line) and the new parametrization \eqref{new_parametrization} proposed in this work (dashed line). For each model the curve is shown only over its physical domain $k_0\le k_{\mathrm{Fn}}\le k_2$, and all panels share common axes. The new form reproduces the shape and peak location of the traditional curves with only minor deviations, which are negligible for practical neutron–star cooling calculations.}
    \label{n3p2_new_fit}
\end{figure*}

Thus, the pairing gap $\Delta$ is one of essential ingredients that determines the impact of the PBF processes in neutron star cooling. The following parametrization has been widely used to represent the density dependence of the pairing gap \cite{Kaminker_2001_parametrization}:
\begin{align}
    \Delta(\kfx, T=0) = \Delta_0  \dfrac{(\kfx - k_0)^2}{(\kfx - k_0)^2+k_1}\dfrac{(\kfx-k_2)^2}{(\kfx-k_2)^2+k_3},\label{traditional_parametrization}
\end{align}
where $\kfx$ is the Fermi wave number for species of baryons specified by `$\mathrm{x}$' ($\in\{\text{n, p}\}$) and $\Delta_0,\,k_0,\, k_1,\, k_2,\,$ and $k_3$ are fitting parameters. This parametrization does not explicitly incorporate the maximum value of the pairing gap, as $\Delta_0$ does not correspond to the gap maximum, which is essential for determining the critical temperature for the onset of superfluidity. For instance, we show in Fig.~\ref{similar_gaps} three gap functions with completely different parameter sets, yet providing very similar results. It is now apparent that even the parameter $\Delta_0$, which appears as if it represents magnitude of the gap function, has no physical meaning. From an optimization perspective, this feature is undesirable, since parameter convergence during optimization does not guarantee convergence to a unique gap model. Consequently, the parametrization in Eq.~(\ref{traditional_parametrization}) is unsuitable for automated parameter optimizations.

To cure this drawback of the traditional gap function \eqref{traditional_parametrization}, here we propose a new parametrization tailored for parameter optimizations:
\begin{align}
     &\Delta(\kfx, T=0) \,= \nonumber\\&\dfrac{\Delta_{\mathrm{max}}(\kfx - k_0)^2 (\kfx - k_2)^2}{(\kfx - k_0)^2 (\kfx - k_2)^2 +w^{-1}(\kfx - k_{\mathrm{max}})^2(1+\alpha(\kfx - k_{\mathrm{max}}))}, \label{new_parametrization}\nonumber\\[1mm]
\end{align}
where $\Deltamax$ now has direct physical meaning of the maximum pairing gap at the Fermi wave number $\kmax$. The parameters $k_0$ and $k_2$ represent the left and right edges of the gap model, respectively. The parameters $w$ and $\alpha$ in the denominator control the width and asymmetry of the gap function. Note that the gap model is defined over the interval $k_0 \leq k \leq k_2$ with boundary conditions $\Delta(k=k_0) = \Delta(k=k_2)=0$, consistent with their definition in Eq.~(\ref{traditional_parametrization}).

To check the validity of our new parametrization, we show in Fig.~\ref{n3p2_new_fit} comparisons of representative neutron $^3\text{P}_2$ pairing gap functions described by the traditional parametrization of Eq.~\eqref{traditional_parametrization} (black solid line) and the new one, Eq.~\eqref{new_parametrization} (red dashed line). As can be seen from the figure, our new parametrization \eqref{new_parametrization} reproduces the eight commonly used neutron $^3\text{P}_2$ pairing gap models fairly well. In Table~\ref{models_and_new_parameters}, we provide the six parameters of the new gap function \eqref{new_parametrization}, fitted to widely-used existing neutron $^1\text{S}_0$, neutron $^3\text{P}_2$, and proton $^1\text{S}_0$ pairing gap models.

\begin{table}[t]
\caption{The parameters of the new pairing gap function \eqref{new_parametrization}. 
For each commonly used gap model listed in Table~II of Ref.~\cite{Ho_2015}, we keep the domain endpoints $k_0$ and $k_2$ identical to Ref.~\cite{Ho_2015} and determine $(\Delta_{\max},\,k_{\max},\,w,\,\alpha)$ by least–squares fits so that the new form reproduces the traditional parametrization \eqref{traditional_parametrization} over $k_0\le k_F\le k_2$. 
Here $\Delta_{\max}$ is the peak value attained at $k_{\max}$, $w$ controls the width, and $\alpha$ the asymmetry. 
The last column lists the original references for each model.}\vspace{3mm}
    \centering
    \begin{tabular}{lccccccc} 
    \hline \hline
        \begin{tabular}{c}
             Gap  \\
             model 
        \end{tabular}  & \begin{tabular}{c}
             $\Delta_{\text{max}}$  \\
             (MeV) 
        \end{tabular} & \begin{tabular}{c}
             $k_{0}$  \\
             (fm$^{-1}$) 
        \end{tabular} & \begin{tabular}{c}
             $k_{2}$  \\
             (fm$^{-1}$) 
        \end{tabular} & \begin{tabular}{c}
             $k_{\text{max}}$  \\
             (fm$^{-1}$) 
        \end{tabular} & \begin{tabular}{c}
             $w$  \\
             (fm$^{2}$) 
        \end{tabular} & \begin{tabular}{c}
             $\alpha$  \\
             (fm) 
        \end{tabular} & \hspace{1mm}Ref. \\
    \hline
    \multicolumn{8}{c}{Neutron singlet (ns)} \\
    AWP2 & 1.3922 & 0.2 & 1.7 & 0.9197 & 0.9278 & 0.0755 & \cite{Ainsworth_1989_ns_AWP2_AWP3} \\
    AWP3 & 1.1500 & 0.2 & 1.4 & 0.8000 & 1.4883 & 0.0000 & \cite{Ainsworth_1989_ns_AWP2_AWP3} \\
    CCDK & 0.8449 & 0.18 & 1.08 & 0.6539 & 2.7410 & -0.1892 & \cite{Chen_1993_ns_CCDK}\\
    CLS & 1.6851 & 0.18 & 1.3 & 0.8044 & 5.8332 & -0.1878 & \cite{Cao_2006_ns_CLS_1, Gandolfi_2009_ns_CLS_2} \\
    GIPSF & 2.0951 & 0.18 & 1.2 & 0.5749 & 2.2783 & 0.5340 & \cite{Gandolfi_2009_ns_CLS_2, Gandolfi_2008_ns_GIPSF} \\
    MSH & 1.7049 & 0.18 & 1.4 & 0.7215 & 3.7356 & 0.1754 & \cite{Gandolfi_2009_ns_CLS_2, Margueron_2008_ns_MSH} \\
    SCLBL & 0.9830 & 0.35 & 1.67 & 1.2834 & 1.4988 & -0.8331 & \cite{Schulze_1996_ns_SCLBL}\\
    SFB & 0.8099 & 0.1 & 1.55 & 0.8482 & 1.0271 & -0.0674 & \cite{Schwenk_2003_SFB}\\
    WAP & 0.9049 & 0.15 & 1.4 & 0.7750 & 1.4026 & 0.0000 & \cite{Schwenk_2003_SFB, wambach_1993_ns_WAP} \\
    \hline
    \multicolumn{8}{c}{Proton singlet (ps)} \\
    AO & 0.3660 & 0.15 & 1.05 & 0.5107 & 2.5608 & 0.6299 & \cite{amundsen_1985_ps_AO_1, Elgaroy_1996_ccdk} \\
    BCLL & 0.8145 & 0.05 & 1.05 & 0.4884 & 3.2606 & 0.2595 & \cite{baldo_1992_ps_BCLL,Elgaroy_1996_ccdk} \\
    BS & 0.7108 & 0.0 & 0.8 & 0.5201 & 3.2569 & -1.0000 & \cite{baldo_2007_ps_BS} \\
    CCDK & 1.0074 & 0.0 & 1.4 & 0.7029 & 1.2855 & -0.1949 &\cite{Chen_1993_ns_CCDK, Elgaroy_1996_ccdk} \\
    CCYms & 0.7801 & 0.0 & 1.1 & 0.6308 & 1.7309 & -0.3891 & \cite{chao_1972_ps_CCY} \\
    CCYps & 0.6678 & 0.0 & 0.95 & 0.5595 & 2.3098 & -0.5439 & \cite{chao_1972_ps_CCY} \\
    EEHO & 0.8807 & 0.0 & 1.2 & 0.6352 & 1.5354 & -0.1174 & \cite{Elgaroy_1996_ccdk} \\
    EEHOr & 1.0076 & 0.0 & 1.1 & 0.6227 & 1.7456 & -0.3567 & \cite{elgaroy_1996_ps_EEHOr} \\
    T & 0.4932 & 0.15 & 1.2 & 0.6685 & 2.0057 & 0.0376 & \cite{Takatsuka_1973_ps_T} \\
    \hline
    \multicolumn{8}{c}{Neutron triplet (nt)} \\
    AO & 0.5887 & 1.2 & 3.3 & 1.9667 & 0.5015 & 0.3309 & \cite{amundsen_1985_nt_AO} \\
    BEEHS & 0.2794 & 1.0 & 3.2 & 2.1818 & 0.9107 & -0.0664 & \cite{baldo_1998_nt_BEEHS} \\
    EEHO & 0.3498 & 1.28 & 2.37 & 1.9664 & 5.5151 & -0.4666 & \cite{Elgaroy_1996_nt_EEHO} \\
    EEHOr & 0.1198 & 1.2 & 1.6 & 1.4356 & 22.3060 & -0.9344 & \cite{elgaroy_1996_ps_EEHOr} \\
    SYHHP & 0.6354 & 2.08 & 2.7 & 2.4450 & 12.1916 & -0.5709 & \cite{shternin_2011_nt_SYHHP} \\
    T & 0.6324 & 1.55 & 2.35 & 1.9295 & 5.5104 & 0.1312 & \cite{amundsen_1985_nt_AO, Takatsuka_1972_nt_T} \\
    TTav & 0.3970 & 1.1 & 2.92 & 1.8511 & 0.6460 & 0.2445 & \cite{Takatsuka_2004_nt_TTav_TToa} \\
    TToa & 0.4486 & 1.1 & 3.2 & 1.9712 & 0.5171 & 0.1956 & \cite{Takatsuka_2004_nt_TTav_TToa} \\
    \hline \hline
    \end{tabular}
    \label{models_and_new_parameters}
\end{table}

Since the new parametrization is not constrained to the conventional bell-shaped gap function, it can generate various functional forms. For our optimization purpose, it is thus necessary to filter out extremely non-physical models. The most problematic case of extreme gap function forms occurs when the function maximum exists near either an end of the function domain---that is, when $k_{\text{max}}$ is close to either the left boundary $k_0$ or the right boundary $k_2$. Since $k_{\text{max}}$ takes values between $k_0$ and $k_2$, it can be expressed using parameter $\beta$ as follows:
\begin{align}
k_{\text{max}} = (1-\beta) k_0 + \beta k_2\quad (0 < \beta < 1). \label{k_max_range}
\end{align}
The position of $k_{\text{max}}$ between $k_0$ and $k_2$ varies according to the value of $\beta$, and as $\beta$ approaches unity, $k_{\text{max}}$ approaches $k_2$. We find that the existing models of the neutron $^3\mathrm{P}_2$ of current interest distribute in a range of $0.365 \leq \beta \leq 0.63$. Based on this observation, we restrict the optimization range to $0.35 \leq \beta \leq 0.65$ to exclude gap functions with extremely non-physical shapes. (For details of this analysis, see Appendix~\ref{appendix:a}.)

Furthermore, as is evident from Eq.~(\ref{new_parametrization}), unlike conventional parametrization (\ref{traditional_parametrization}), cases exist where the gap function diverges within the domain $k_0 \leq k \leq k_2$. This behavior is determined by the asymmetry parameter $\alpha$, and it is necessary to establish the range of $\alpha$ values that ensure the gap function remains finite within the domain $k_0 \leq k \leq k_2$ for physical validity. One can readily suspect such cases where the denominator becomes zero. For $k_0 < k < k_2$ with $k \neq k_{\text{max}}$, since $(k-k_0)^2(k_2-k)^2 > 0$, $w > 0$, and $(k-k_{\text{max}})^2 > 0$, problems arise when $1+\alpha (k- k_{\text{max}}) < 0$. Therefore, by imposing the condition $1+\alpha (k- k_{\text{max}}) > 0$ and considering the cases $k > k_{\text{max}}$ and $k < k_{\text{max}}$ separately, we obtain:
\begin{align}
-\dfrac{1}{k_2 - k_{\text{max}}} < \alpha < -\dfrac{1}{k_0 - k_{\text{max}}},\label{alpha_condition_one}
\end{align}
or, substituting Eq.~(\ref{k_max_range}) into this expression, we have:
\begin{align}
-\dfrac{1}{(1-\beta) (k_2-k_0)} < \alpha < \dfrac{1}{\beta(k_2 - k_0)} \label{al_range}.
\end{align}
By investigating the $\alpha$ distribution for existing models, we have confirmed that most $\alpha$ values cluster near the center of the mathematically allowed range. To allow broader exploration in our optimization, we trim this range by excluding the outermost 10\% at each end [\textit{i.e.}, we use 80\% of the interval implied by Eq.~(\ref{al_range})]. (These are also discussed in Appendix~\ref{appendix:a}.) The full numerical bounds for all parameters used in the search are summarized later in Table~\ref{tab:parameter_bounds}.

\begin{table}[t]
\centering
\caption{Parameter bounds used in the TPE optimization. $\Delta_{\max}$ caps the peak height;
$k_0$ and $k_2$ delimit the pairing-gap domain; $k_{\max}$ sets the peak location;
$w$ controls the width; and $\alpha$ skews the shape, with its range restricted to the
central 80\% of the analytically allowed interval defined by Eq.~(\ref{al_range}).
}\vspace{3mm}
\begin{tabular}{lcc}
\hline
Parameter & Min & Max \\
\hline
$\Delta_{\max}$ & 0.10 & 1.50 \\
$k_0$            & 0.90 & 2.50 \\
$k_2$            & 1.50 & 3.50 \\
$k_{\max}$       & 1.00 & 3.00 \\
$w$              & 0.2 & 100 \\
$\alpha$            & \multicolumn{2}{c}{Eq.~(\ref{al_range})} \\
\hline
\end{tabular}\vspace{-3mm}
\label{tab:parameter_bounds}
\end{table}

\subsection{Neutron star model}

The internal structure of neutron stars depends critically on the equation of state (EoS) of dense matter. Over the past decades, diverse approaches have been advanced: variational many-body and Brueckner–Hartree–Fock calculations (\textit{e.g.}, APR) \cite{Akmal_1998_APR}, Skyrme energy density functionals and unified crust–core EoS (\textit{e.g.}, SLy) \cite{DouchinHaensel_2001_SLy}, relativistic mean-field models (\textit{e.g.}, GM1, DD2) \cite{Glendenning_1991_GM1,Typel_2010_DD2}; In this work we adopt the BSk24 equation of state (EoS) \cite{Pearson_2018_bsk24}, a modern Brussels--Skyrme functional widely used in neutron star cooling studies \cite{Leinson_2022_hybrid_cooling}. The present study is not intended to assess the dependence of the results on the choice of the EoS, but to explore, within a fixed setup, the relationship between the PBF efficiency factor $q$ and the neutron ${}^{3}\mathrm{P}_{2}$ pairing-gap shape. A systematic investigation of EoS dependence is left to a follow-up Bayesian inference study \cite{Bayesian_Cooling}.

Assuming spherically-symmetric neutron stars, the internal structure can be calculated by solving the Tolman-Oppenheimer-Volkoff (TOV) equation \cite{TOV_1939} with the determined EoS, allowing us to derive the neutron star mass and radius for a given central density. Since BSk24 is a unified EoS, we construct EoS tables for the outer core and inner crust regions by referencing the publicly available Fortran77 fitting program (\texttt{bskfit18.f}) \cite{Pearson_2018_bsk24, BSk_subroutine_website}, and create the outer crust EoS table following Table 4 in Ref.~\cite{Pearson_2018_bsk24}. Using our TOV solver developed in Fortran90, we solve the following equations:
\begin{align}
& \frac{\dd m}{\dd r}=4 \pi r^2 \rho, \\
& \frac{\dd\Phi}{\dd r}=\frac{G m c^2+4 \pi G r^3 P}{c^4 r^2\left(1-\frac{2 G m}{c^2 r}\right)}, \\
& \frac{\dd P}{\dd r}=-\frac{\left(\rho+\frac{P}{c^2}\right)\left(G m+\frac{4 \pi G r^3 P}{c^2}\right)}{r^2\left(1-\frac{2 G m}{c^2 r}\right)}, \\
& \frac{\dd a}{\dd r}=\frac{4 \pi r^2 \nb}{\sqrt{1-\frac{2 G m}{c^2 r}}},
\end{align}
where $\nb$ is the number density of baryons, $c$ is the speed of light in vacuum, $a(r)$ denotes the enclosed baryon number within radius $r$, and $G$ is the gravitational constant. $\Phi=\phi/c^2$ is the metric function, where $\phi$ is the gravitational potential in Newtonian mechanics, which satisfies the boundary condition that the metric inside the star must match the exterior (vacuum) Schwarzschild metric at the stellar radius $r=R$:
\begin{align}
e^{\Phi(R)} = \sqrt{1-\frac{2GM}{c^2R}},
\end{align}
where $M=M_R$ is the gravitational mass of the star. These equations are solved by integrating outward from $r=0$, with $\rho$ and $P$ for each $\nb$ maintained according to the EoS, until $P=0$ (the neutron star surface) is reached. Based on these calculations, we obtain structure profiles for each mass, which will be utilized in subsequent neutron star cooling calculations.

Since the internal structure of neutron stars varies with mass, accurate mass determination is considered essential for successful cooling modeling. Unfortunately, the precise mass of the Cas~A NS has not yet been determined. Recent X-ray spectral analysis of Cas~A suggests that the neutron star mass is approximately $(1.55\pm 0.25)M_\odot$ \cite{Shternin_2022_1.55_0.25}. In the present work, we do not attempt to constrain the neutron star mass. Instead, we
adopt a fixed canonical mass of $M_{\mathrm{NS}} = 1.4\,M_\odot$, which lies below the direct-Urca threshold of $1.595\,M_\odot$ for the BSk24 equation of state \cite{pearson_2018_bsk24_urca_threshold} and is therefore consistent with a PBF-dominated cooling scenario. The effects of mass variation, together with statistically robust constraints on $q$ and on the neutron ${}^{3}\mathrm{P}_{2}$ gap parameters, will be addressed in a dedicated follow-up Bayesian inference study \cite{Bayesian_Cooling}.

\subsection{Neutron star cooling}

\subsubsection{Basic equations and assumptions}

The thermal evolution (or cooling) of neutron stars can be described by the following equations (see, \textit{e.g.}, Ref.~\cite{Thorne_1977, Yakovlev_2004_cooling_equation}):
\begin{align}
\frac{{\partial}\left(L e^{2 \Phi}\right)}{{\partial} r} &=-\frac{4 \pi r^2 e^{\Phi}}{\sqrt{1-2 G m / c^2 r}}\left(C_V \frac{{\partial} T}{{\partial} t}+e^{\Phi}\left(Q_\nu-Q_\text{h}\right)\right), \label{Eq:thermal_evolution_1}\\ 
\frac{{\partial}\left(T e^{\Phi}\right)}{{\partial} r}&=-\frac{1}{\lambda} \cdot \frac{L e^{\Phi}}{4 \pi r^2 \sqrt{1-2 G m / c^2 r}}, \label{Eq:thermal_evolution_2}
\end{align}
where $L$ and $T$ denote luminosity and temperature, respectively. The upper (lower) equation corresponds to the energy balance (transport). In these equations, $\lambda$ is the thermal conductivity, $C_V$ is the heat capacity per unit volume, and $Q_\nu$ and $Q_\text{h}$ are the neutrino emissivity and heating rate, respectively, both with units of energy per unit volume per unit time. $\Phi$ is the metric function that can be obtained by solving the TOV equation. In this study, we do not consider any heating sources; therefore, $Q_\text{h}=0$. Note that we assume that neutrinos completely escape from the neutron star.

In this study, we employ the barotropic EoS approximation, which is one of the most commonly used methods for solving the thermal evolution equations (see, \textit{e.g.}, Refs.~\cite{Gnedin_2001, Page_2016_NSCool}). This approximation recognizes that matter is strongly degenerate in the sufficiently-high density regions of neutron star interiors, allowing separate treatment of the internal structure and thermal structure in neutron star cooling calculations \cite{Potekhin_2018_magnetic}. Therefore, we specify a reference boundary mass density $\rho_\text{b}$ (the most widely accepted value is $\rho_\text{b} = 10^{10}\,\mathrm{g/cm^3}$ \cite{Gudmundsson_1983_envelope}) and solve the thermal evolution equations under the barotropic EoS assumption for $\rho > \rho_\text{b}$. The boundary radius $r=r_\text{b}$ [\textit{i.e.} $\rho(r_\text{b})=\rho_\text{b}$] corresponds to the outer boundary in the calculation.

The envelope of the neutron star, existing in the lower-density region $\rho < \rho_\text{b}$, is the region with the largest temperature gradient and is treated using a function called the $T_\text{s}$--$T_\text{b}$ relation. This corresponds to a functional fit that provides the relationship between the actual surface temperature $T_\text{s}$ and the temperature $T_\text{b}$ at the bottom of the envelope. The calculation separates the region between the envelope bottom (at $\rho \simeq 10^{10}\,\mathrm{g/cm^3}$, or lower densities such as $\rho \simeq 10^{8}\,\mathrm{g/cm^3}$ for shorter-timescale cooling descriptions) and the surface from the cooling calculation in higher-density regions \cite{Pons_2019}.

The surface temperature $T_\text{s}$ is related to the photon luminosity $L_{\gamma}$ as follows:
\begin{align}
  L_{\gamma} = \sigma_{\mathrm{SB}}\int T_s^4 \dd\Sigma = 4\pi R^2 \sigma_{\mathrm{SB}}T_{\mathrm{eff}}^4,
\end{align}
where $\sigma_{\mathrm{SB}}$ is the Stefan-Boltzmann constant, $\dd\Sigma$ is the surface element. The so-called effective temperature $T_{\mathrm{eff}}$ is introduced since the distribution of the surface temperature $T_\text{s}$ over the neutron star's surface can be non-uniform due to magnetic fields, atmospheric structure, etc. However, under the assumption of a spherical symmetry without magnetic field, it coincides with the surface temperature, \textit{i.e.} $T_{\mathrm{eff}} = T_\text{s}$. Note that the quantities $L_{\gamma},T_{\mathrm{eff}}$, and $T_\text{s}$ refer to a local reference frame at the surface of the neutron star. The quantities observed by a distant observer are redshifted as follows (see, \textit{e.g.}, Ref.~\cite{Thorne_1977}):
\begin{align}
    L_{\gamma}^\infty &= L_{\gamma}\left( 1-\dfrac{2GM}{c^2R} \right) = 4\pi\sigma_{\mathrm{SB}}(T_{\mathrm{eff}}^\infty)^4R_{\infty}^2,\\
    T_{\mathrm{eff}}^\infty &= T_{\mathrm{eff}}\sqrt{ 1-\dfrac{2GM}{c^2R}}, \\
    R_\infty &= \dfrac{R}{\sqrt{ 1-\dfrac{2GM}{c^2R}}}.
\end{align}

In this study, we use the $T_{\mathrm{s}}$--$T_{\mathrm{b}}$ relation presented in Ref.~\cite{Potekhin_1997_accreted_envelopes}. As a specific and fixed setup, we adopt a thin carbon envelope with mass $\Delta M = 10^{-15}\,M_\odot$ on top of an iron envelope, following Ref.~\cite{Ho_2015}, where this configuration was identified as part of the best-fit solution from spectral analyses of the Cas~A neutron star. Since the envelope composition and mass can have a significant impact on neutron star cooling and remain subject to substantial uncertainty, a systematic treatment of $\Delta M$ as a free parameter is deferred to a dedicated follow-up Bayesian inference study \cite{Bayesian_Cooling}.

\subsubsection{Implementation of a computational code for neutron star cooling}

For the numerical computation of neutron star cooling, 
Eqs.~\eqref{Eq:thermal_evolution_1} and \eqref{Eq:thermal_evolution_2} 
must be discretized and solved. 
First, to simplify the form of these equations, 
we introduce the redshifted temperature and luminosity,
\[
\mathcal{T} \equiv e^{\Phi} T, \quad 
\mathcal{L} \equiv e^{2\Phi} L,
\]
and define the baryon number coordinate
\[
{\dd}a = \frac{4\pi r^2 n_\mathrm{B}\, \dd r}{\sqrt{1 - 2Gm/c^2r}}.
\]
With these definitions, 
Eqs.~\eqref{Eq:thermal_evolution_1} and \eqref{Eq:thermal_evolution_2} 
can be rewritten as
\begin{align}
\frac{{\partial} \mathcal{T}}{{\partial} t} &= 
F\left( \mathcal{T}, \frac{{\partial}\mathcal{L}}{{\partial} a} \right)
= -e^{2\Phi} \frac{Q_\nu - Q_\mathrm{h}}{C_V}
-\frac{n_\mathrm{B}}{C_V} \frac{{\partial} \mathcal{L}}{{\partial} a}, \label{Eq:thermal_eq_simplified_1}\\
\mathcal{L} &= 
G\left( \mathcal{T}, \frac{{\partial}\mathcal{T}}{{\partial} a} \right)
= -\lambda (4\pi r^2)^2 n_\mathrm{B} e^{\Phi} 
\frac{{\partial} \mathcal{T}}{{\partial} a}.\label{Eq:thermal_eq_simplified_2}
\end{align}

{To solve the thermal evolution equations numerically, we adopt the standard finite-difference
scheme and implicit time-stepping strategy as implemented in the publicly available
\texttt{NSCool} code \cite{Page_2016_NSCool}. In particular, we follow the same discretization
of the stellar interior, the use of redshifted variables, and the even--odd grid structure
for temperature and luminosity, together with the Newton-Raphson iteration scheme and
adaptive time stepping (see, \textit{e.g.}, Refs.~\cite{Page_2016_NSCool,NSCool_lecture_1}).}

{In addition, since the present work explores only the neutron ${}^3\mathrm{P}_2$ pairing-gap parameters, the thermal evolution during the early stage---before the onset of neutron ${}^3\mathrm{P}_2$ superfluidity---is identical across different ${}^3\mathrm{P}_2$ gap models within our setup. We exploit this fact to accelerate the optimization as follows. We first compute and store a reference cooling profile in which neutron ${}^3\mathrm{P}_2$ superfluidity is absent. For each trial set of ${}^3\mathrm{P}_2$ gap parameters, we then evaluate the corresponding critical temperature profile and identify the first epoch at which the temperature at one or more grid points drops below the local critical temperature. We restart the full cooling evolution a few time steps prior to this epoch, thereby avoiding redundant integration over the early, ${}^3\mathrm{P}_2$ gap-insensitive phase. Furthermore, since the Cas~A~NS data correspond to ages of $\sim 300$--$400$~yr, we calculate cooling curves only up to $500$~yr for the optimization runs. These procedures reduce the runtime to $\sim 0.5$~s per cooling curve\footnote{
All computations were performed on a local workstation running macOS~15.6.1 (arm64),
equipped with an Apple~M4 CPU and 16~GB of unified memory.
}.
}

\subsubsection{Cas~A NS observational data}

We use the observational data of the Cas~A NS measured with \textit{Chandra} ACIS-S over the past 20 years, as reported in Ref.~\cite{Shternin_2022_1.55_0.25}. The dataset includes both GRADED observations (14 epochs from 2000 to 2019) and FAINT observations (4 epochs from 2006 to 2020), for a total of 18 epochs with comprehensive temporal coverage. {Throughout this work, we adopt the variable column-density
($N_{\rm H}$) series from the joint ACIS analysis. This choice is treated as a fixed
observational setup in the present exploratory study. The impact of alternative
assumptions for $N_{\rm H}$, including the fixed-$N_{\rm H}$ series, will be examined as
part of a dedicated follow-up Bayesian inference analysis \cite{Bayesian_Cooling}.}

For our cooling simulations, we adopt the canonical neutron star mass of $M_{\rm NS}=1.4\,M_\odot$, below the direct-Urca threshold of $1.595\,M_\odot$ for the BSk24 equation of state, ensuring consistency with the PBF paradigm. The corresponding radius is $R=12.58$\,km. To compare theoretical cooling curves with the observations, we rescale the reported effective temperatures to our stellar model using $T_{\rm eff}^4 R^2=\mathrm{const.}$, which in logarithmic form reads
\begin{equation}
\log_{10} T_{\rm eff}^{\rm (corr)}
=
\log_{10} T_{\rm eff}^{\rm (data)}
+\frac{1}{2}\log_{10}\!\left(\frac{R_{\rm data}}{R_{\rm model}(M)}\right),
\end{equation}
{where $R_{\rm data}=13.7$\,km denotes the analysis radius adopted in Ref.~\cite{Shternin_2022_1.55_0.25} for the spectral fitting in the $N_{\rm H}$-variable setup (see Tables~1 and 2 therein).}

The Cas~A NS dataset is among the most tightly constrained cases of real-time neutron star cooling and thus serves as an exacting testbed for our models of neutron $^{3}\mathrm{P}_{2}$ superfluidity. The rapid temperature decline in this young object ($\sim340$\,yr) strongly points to enhanced neutrino emission associated with the onset of neutron Cooper-pair formation in the core.

{\subsection{Hybrid optimization strategy}}

The optimization of our six-parameter neutron superfluid gap model \eqref{new_parametrization}
poses a substantial computational task because the parameter space is high-dimensional and
must be explored through repeated cooling curve evaluations.
Our approach integrates our Fortran90-based cooling simulation code with Python-based optimization analysis using the Optuna \cite{Akiba_2019_Optuna} framework, creating a seamless computational pipeline for parameter exploration.

Traditional optimization methods such as grid search or random search are computationally prohibitive for this problem. A grid search with even modest resolution (\textit{e.g.}, 10 points per parameter) would require $10^6$ evaluations, while random search lacks the efficiency to converge within reasonable computational limits. To address this challenge, we employ the Tree-structured Parzen Estimator (TPE) algorithm \cite{Bergstra_2011_TPE}, a sequential model-based optimization technique that efficiently navigates high-dimensional parameter spaces.

{Building on this approach, we adopt a hybrid optimization strategy that combines the strengths of global
and local search methods. The TPE algorithm is particularly effective for global exploration
of high-dimensional and multimodal parameter spaces, but it is not designed to achieve precise
local convergence once a promising region has been identified. To refine candidate solutions
obtained from the TPE search, we therefore apply a direct minimization algorithm based on the
Nelder-Mead simplex method \cite{Nelder_1965_NM}. While Nelder-Mead depends on the choice of initial conditions and
can converge to local minima, its application after the TPE-based global exploration mitigates
these limitations. In this complementary framework, TPE identifies promising regions of parameter
space, and Nelder-Mead provides accurate local refinement of the optimal solutions.}

{The optimization is performed within the parameter bounds summarized in
Table~\ref{tab:parameter_bounds} in Sec.~\ref{sf_gap_model}. For each trial, parameter values
are sampled and first checked against physical constraints, including
$0.35 \leq \beta \leq 0.65$ for the position of the gap maximum and the analytically derived
bounds for $\alpha$ given in Eq.~(\ref{al_range}). Only parameter sets that satisfy these
conditions proceed to the cooling calculation.}

{We first perform a global exploration of the parameter space using the TPE algorithm, running $10{,}000$ trials. From these TPE trials, we select a small subset of the best-performing candidates based on the $\chi^2$ statistic. Starting
from the top-ranked solutions, we then apply the Nelder-Mead simplex method to further refine the parameter values through local optimization. This two-stage procedure combines efficient global exploration with accurate local minimization, thereby enabling a systematic identification of neutron superfluid gap models that best reproduce the observed cooling
behavior of the Cas~A~NS.}

\section{Results and Discussion}\label{Sec:Results}

\subsection{Computational setup}

In this study, we employ our newly developed computational code for neutron star cooling written in Fortran90. The cooling code solves the energy transport and energy balance equations---the so-called thermal evolution equations [Eqs.~\eqref{Eq:thermal_evolution_1} and \eqref{Eq:thermal_evolution_2}]---within the general relativistic framework. Note that we assume spherically symmetric neutron stars, so that the thermal evolution equations become one-dimensional ones, \textit{i.e.}, only considering the radial direction without angular dependence. The initial temperature is set to $Te^{\Phi}=10^{10}\,\mathrm{K}$. For the nuclear EoS, we use the unified BSk24 model for both the core and crust. 
{For the proton ${}^1\mathrm{S}_0$ pairing gap, we adopt the widely used CCDK model \cite{Elgaroy_1996_ccdk}, which can support neutron stars with a fully superconducting proton core \cite{Ho_2015}. For the neutron ${}^1\mathrm{S}_0$ pairing, we choose the SFB model \cite{Schwenk_2003_SFB}. These choices are treated as a fixed microphysical setup in the present study. While both proton and neutron ${}^1\mathrm{S}_0$ pairing gaps are known to be subject to theoretical uncertainties, a systematic exploration of alternative models or parameterizations (including recent developments for neutron ${}^1\mathrm{S}_0$ pairing \cite{Ding_2016_n1s0_new}) is deferred to a follow-up Bayesian inference analysis \cite{Bayesian_Cooling}.}
Our microphysics largely follows standard practice (of, \textit{e.g.}, Ref.~\cite{Potekhin_2015_review_physical_input}), and we do not include the in-medium modified-Urca enhancement in Ref.~\cite{Shternin_2018_enhanced_murca} because it was derived for non-superfluid matter and is not directly consistent with our PBF-focused, superfluid setup.

\subsection{Cooling curves}

\begin{figure}[t]
    \centering
    \includegraphics[width=1.0\linewidth]{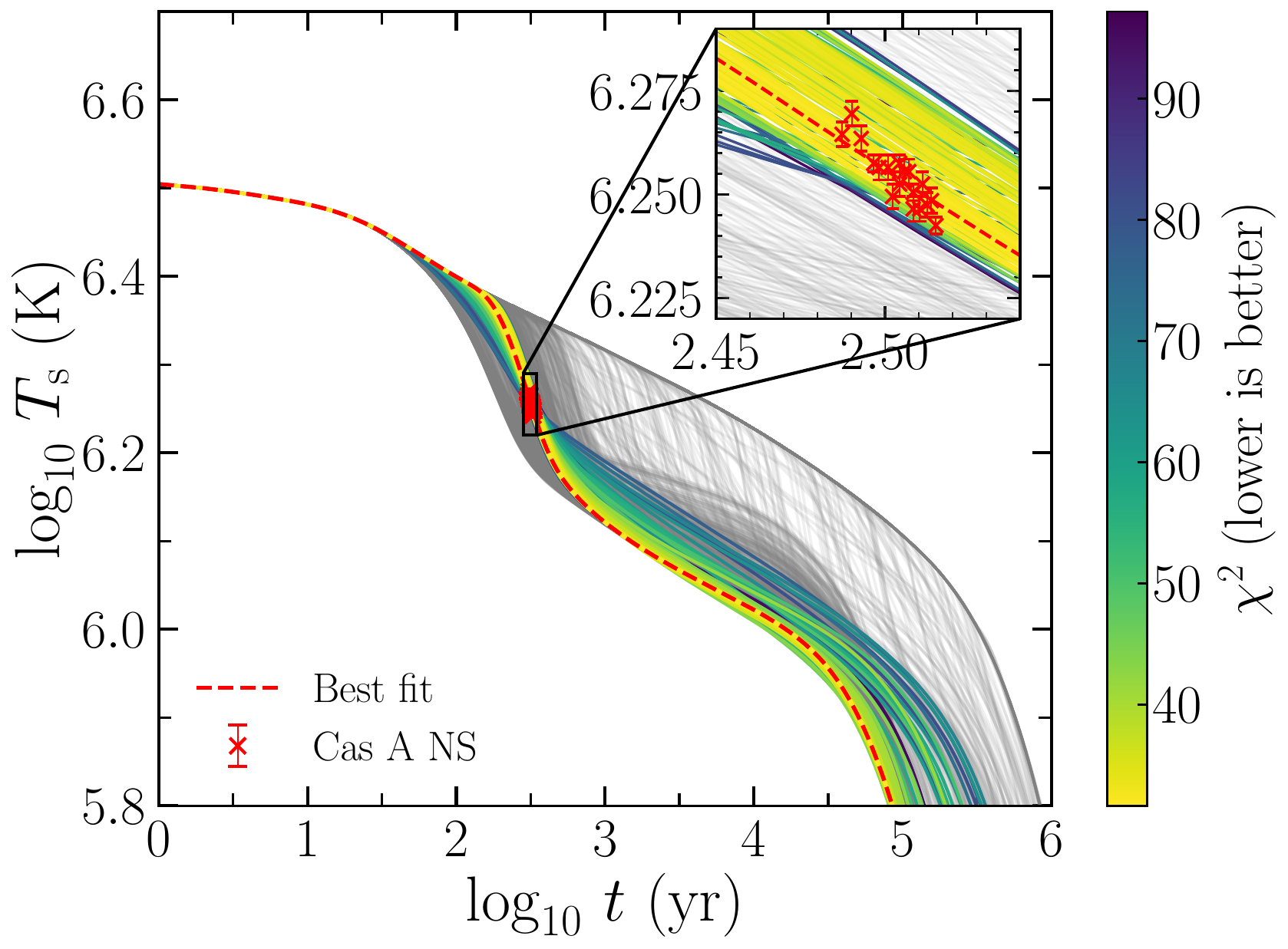}\vspace{-3mm}
   \caption{ Cooling curves obtained from $1{,}000$ trials during the TPE optimization, {shown for
demonstrative purposes.} Each line represents a cooling calculation for a distinct parameter set sampled by TPE and is compared with the Cas~A neutron star data (red data points). The red dashed curve indicates the best-fitting solution used to align the data points for visualization.}
    \label{all_trials_plot}
\end{figure}

\begin{figure*}
    \centering
    \includegraphics[width=1.0\textwidth]{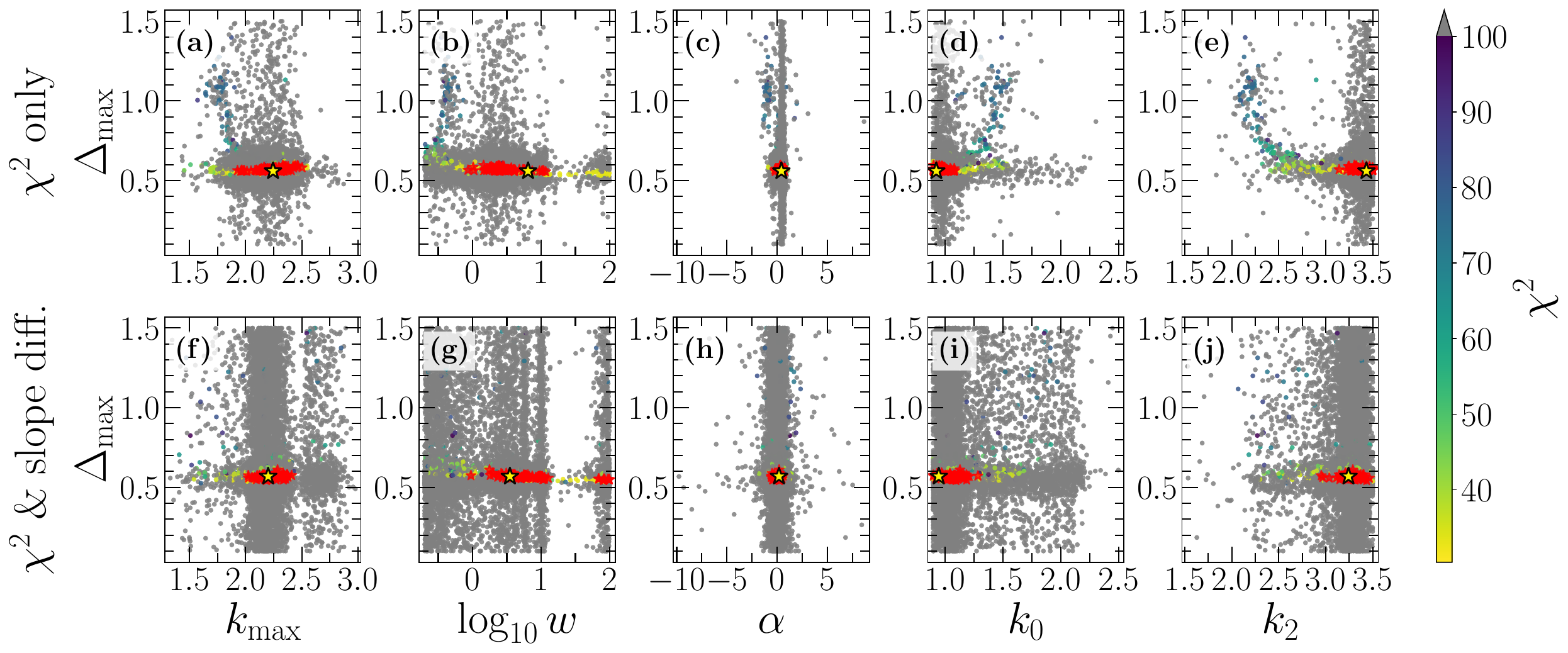}\vspace{-3mm}
    \caption{Parameter-space projections of $\Delta_{\max}$ versus each gap parameter for the single-objective (top row) and multi-objective (bottom row) optimizations. {Each point represents a trial sampled by the TPE algorithm and is colored by the corresponding $\chi^{2}$ value, with brighter colors indicating better agreement with the Cas~A~NS data. Red markers highlight the top 1\% of solutions ranked by $\chi^{2}$ (100 out of 10{,}000). The yellow star denotes the final solution obtained after the hybrid optimization, where the best TPE candidates are further refined using the Nelder-Mead method. For the width parameter, we plot $\log_{10} w$ to account for its wide dynamic range and to improve readability.}}
    \label{parameter_relation_comparison_for_objective_type}
\end{figure*}

To showcase the idea and feasibility of the proposed approach, we show in Fig.~\ref{all_trials_plot} an illustrative subset of 1,000 trials from the optimization procedure, where cooling curves were computed for a wide range of parameter sets sampled by TPE to assess their consistency with the Cas~A NS data. In the figure, the surface temperatures of neutron stars are plotted as functions of their age in a double logarithmic plot. Here we show 1,000 cooling curves associated with different sets of neutron $^3\text{P}_2$ pairing gap parameters. Line colors indicate the $\chi^2$ score, where lighter colors are better, while gray lines correspond to trashy parameter sets with $\chi^2>100$. Red crosses with an error bar show the Cas~A NS observational data, whereas red dashed line corresponds to the best fit result after the optimization. In the inset, cooling curves close to the Cas~A data are exhibited.

From the figure, we observe several distinct features in the cooling behavior obtained from the TPE optimization trials. Since it provides us rich and useful information, let us discuss global behaviors of the cooling curves accumulated during the TPE optimization process.

A considerable number of trials produced cooling curves in which neutron $^3\mathrm{P}_2$ pairing did not occur, and only neutron $^1\mathrm{S}_0$ superfluidity and proton $^1\mathrm{S}_0$ superconductivity were active. These curves correspond to the highest-temperature group of solutions, extending up to $t \sim 10^6$~yr. Such behavior arises because, for these parameter sets, the gap amplitude of neutron $^3\mathrm{P}_2$ was not sufficiently large within the Fermi-momentum range realized in the stellar core, and therefore the pairing transition did not occur within the temperature range shown in the plot.

The small shoulder structure appearing around $t \sim 10^3$--$10^4$~yr is attributed to a relatively early onset of neutron $^3\mathrm{P}_2$ pairing, which leads to an earlier suppression of the neutrino emissivity including that of PBF process compared with other models. 
As a result, the cooling slows down, producing a characteristic flattening of the curve in that period.

The fact that many trial curves cluster around the region where the Cas~A NS is located suggests that the TPE algorithm effectively performs optimization toward this observational constraint. 
This concentration indicates that the parameter space around the best-fit solution is well explored and efficiently sampled.

Although the neutron and proton $^1\mathrm{S}_0$ pairing emerge much earlier than the lower bound of the plotted range ($t=1$~yr), noticeable differences in surface temperature appear only for $1 \lesssim t \lesssim 10$~yr. Since these pairing gaps are fixed to the SFB \cite{Schwenk_2003_SFB} and CCDK \cite{Elgaroy_1996_ccdk} models, respectively, all cooling curves are identical before the appearance of neutron $^3\mathrm{P}_2$ pairing. Therefore, the variation among the final cooling curves originates solely from differences in the neutron $^3\mathrm{P}_2$ gap model.

In the best-fit case, the sharp decline in temperature immediately after $t \gtrsim 100$~yr corresponds to the onset of PBF neutrino emission associated with neutron $^3\mathrm{P}_2$ pairing, which temporarily enhances the total neutrino luminosity. 
After this rapid cooling phase, the slope becomes gentler as the neutrino emissivity from the neutron $^3\mathrm{P}_2$ PBF process decreases with temperature. 
A second steepening occurs around $\log_{10} t \sim 4.5$, marking the transition from the neutrino-emission era to the photon-radiation era, where surface photon radiation dominates the cooling.

In the inset, some cooling curves appear to yield small $\chi^2$ values despite not directly intersecting the plotted Cas~A NS data points. 
This is because the Cas~A NS data are allowed to shift within $\pm 19$~yr relative to each model curve when evaluating the fit. 
The data points displayed in the figure are aligned to the best-fit curve, while in reality the Cas~A NS observational data is assumed to have an uncertainty of $\pm 19$~yr around the true birth epoch of the star.

\subsection{Single-objective ($\chi^2$) vs.\ multi-objective ($\chi^2$\,+\,slope-difference) optimizations}
\label{subsec:opt_compare}

We compare two formulations at fixed $q\simeq 0.19$ and $M_\text{NS}=1.4\,M_{\odot}$: (i) a single-objective optimization that minimizes only the misfit between theory and data, quantified by the chi-squared statistic $\chi^2$; and (ii) a multi-objective optimization that jointly minimizes $\chi^2$ and the absolute difference between the local slope of the theoretical cooling curve and the slope obtained from a linear fit to the data around its temporal midpoint (hereafter we call the latter ``slope diff.'').

For each model we evaluate the misfit using
$$
\chi^2=\sum_i\left[\frac{\log_{10} T_i^{\rm (data)}-\log_{10} T^{\rm (model)}(t_i;t_0)}{\sigma_i}\right]^2,
$$
minimizing over a single nuisance parameter, the age offset $t_0\in[-19,+19]$~yr, to account for the birth-epoch uncertainty. Here $\sigma_i$ are the $1\sigma$ uncertainties of $\log_{10} T_i^{\rm (data)}$. With $N=18$ measurements and one fitted parameter ($t_0$), the nominal degrees of freedom are $\nu=N-1=17$. Unless noted, we report $\chi^2$ only; the corresponding reduced value $\chi^2_\nu\equiv\chi^2/\nu$ (with $\nu=17$) can be obtained by simple rescaling. The neutron $^{3}\mathrm{P}_{2}$ gap parameters are chosen by the global optimization prior to the $\chi^2$ evaluation and are not varied within a given fit, hence they do not enter the degrees-of-freedom count.

To demonstrate the difference between the single- and multi-objective optimizations, we show in Fig.~\ref{parameter_relation_comparison_for_objective_type} the explored regions of parameter space for $\Delta_{\max}$ versus each of the remaining five parameters ($k_\text{max}$, $w$, $\alpha$, $k_0$, and $k_2$) for both optimization types. Point colors indicate the $\chi^2$ score (lighter color is better). Because the optimization is six-dimensional, any 2D projection may place nearby points that are distant in the remaining coordinates; the plots should therefore be interpreted as projections. Red markers denote the top 1\% in $\chi^2$ (100 best trials out of 10,000), {while the yellow star indicates the final solution obtained after further local refinement of the best TPE candidates using the Nelder-Mead method.}

\begin{figure}[t]
    \centering
    \includegraphics[width=1.0\linewidth]{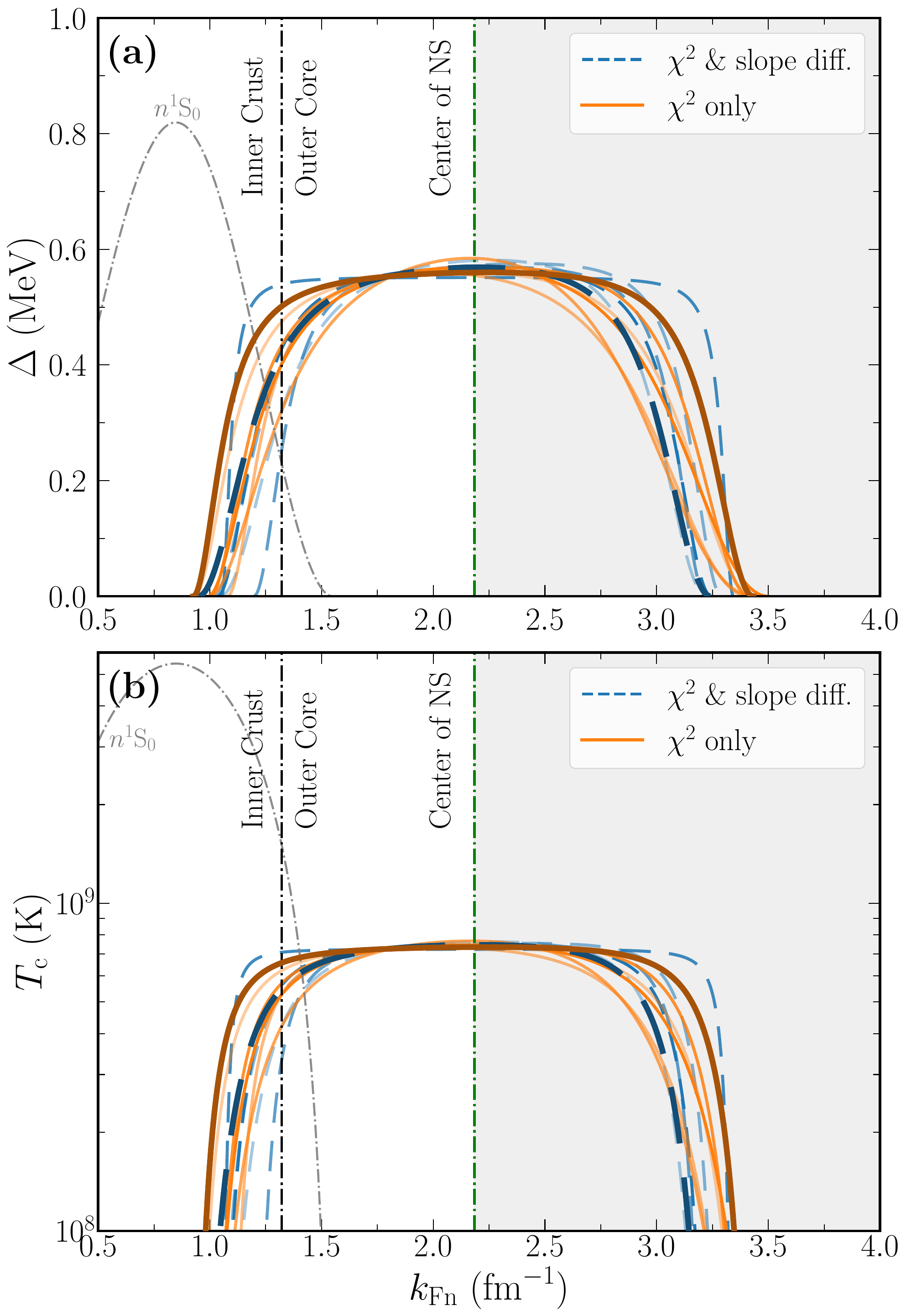}\vspace{-3mm}
    \caption{{Top-5 gap functions $\Delta(k_{\mathrm{Fn}})$ (a) and critical-temperature profiles
$T_{\mathrm{c}}(k_{\mathrm{Fn}})$ (b) obtained from the TPE optimization for each objective
type, with darker curves indicating lower $\chi^{2}$. The final best-fit solution obtained
after the subsequent Nelder-Mead refinement is highlighted by a thicker and darker curve.
The vertical black dashed line marks the inner-crust/outer-core boundary in
$k_{\mathrm{Fn}}$, while the vertical green dashed line marks the central
$k_{\mathrm{Fn}}$. The gray region to the right is not realized inside the star, explaining
the comparatively large uncertainty in $k_{2}$. The gray dashed curve shows the neutron
${}^{1}\mathrm{S}_{0}$ SFB model used as a reference.}}
    \label{objective_type_gap_and_tc}
\end{figure}

In the $\Delta_{\max}$--$k_{\max}$ plane [Figs.~\ref{parameter_relation_comparison_for_objective_type}(a) and \ref{parameter_relation_comparison_for_objective_type}(f)], the multi-objective run concentrates its top-1\% solutions in a narrower band centered near {$k_{\max}\!\approx\!2.25$}, whereas the single-objective run exhibits a broader spread. {This reflects the fact that the multi-objective optimization simultaneously requires both
the overall goodness of fit ($\chi^{2}$) and the cooling slope difference to be small, thereby favoring a more restricted region of parameter space in which both criteria are satisfied.} The same qualitative trend appears in the other $\Delta_{\max}$--(parameter) projections, with one notable exception: for $k_2$ [Figs.~\ref{parameter_relation_comparison_for_objective_type}(e) and \ref{parameter_relation_comparison_for_objective_type}(j)] the top-tier solutions of the multi-objective run are more widely distributed than in the single-objective case. This is natural because $k_2$ sets the right-hand edge of the gap in neutron Fermi momentum; once $k_2$ exceeds the neutron Fermi momentum at the center of the star, further increases in $k_2$ have no observable impact on the cooling physics (see also Fig.~\ref{objective_type_gap_and_tc}(a)).

\begin{figure}[t]
    \centering
    \includegraphics[width=1.0\linewidth]{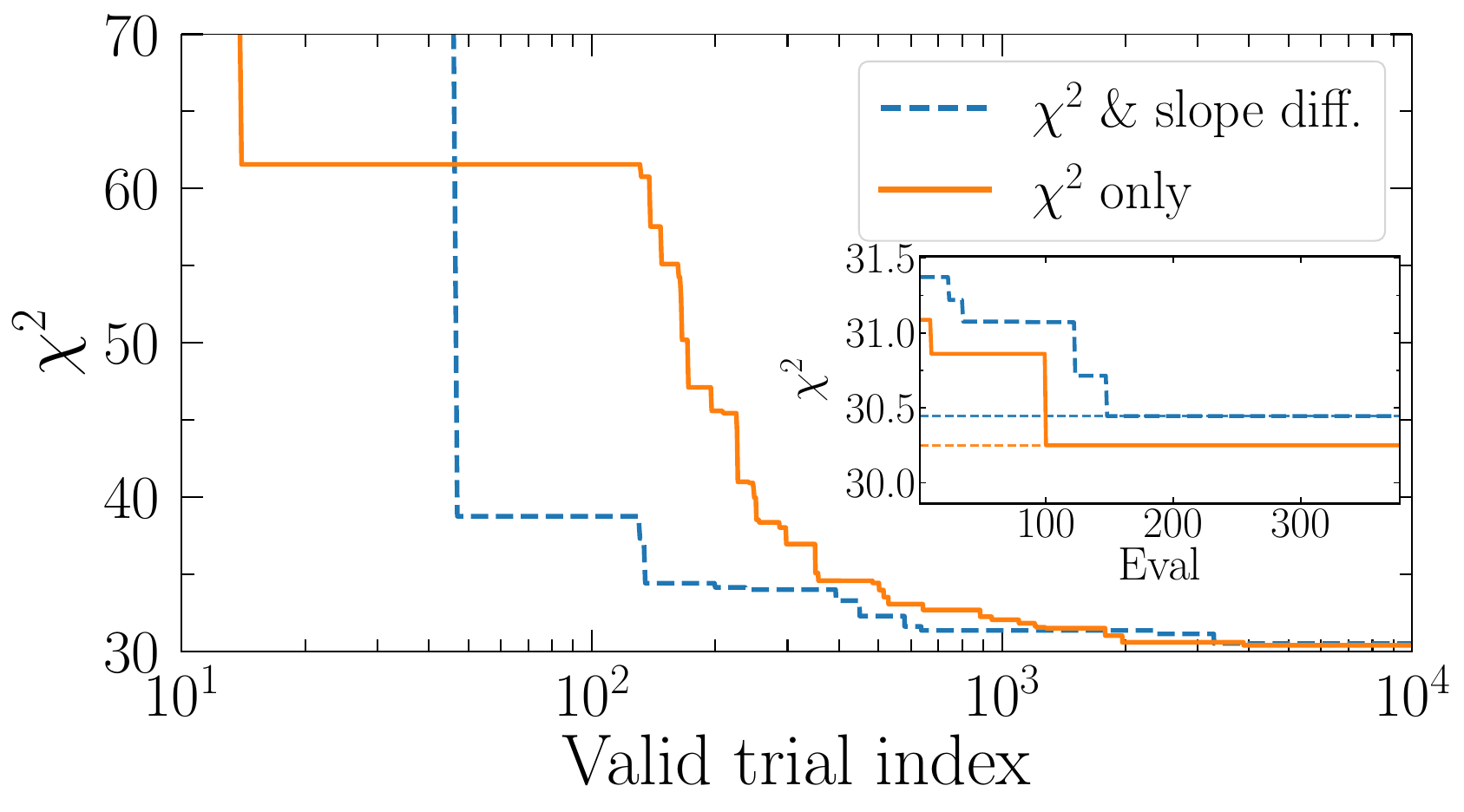}\vspace{-4mm}
  \caption{{Evolution of the best (top-1) $\chi^{2}$ as a function of the number of valid trials for
the single-objective ($\chi^{2}$ only) and multi-objective
($\chi^{2}$+$\,\mathrm{slope\ diff.}$) TPE runs at fixed $q\simeq 0.19$ and
$M_{\rm NS}=1.4\,M_\odot$. The horizontal axis is shown on a logarithmic scale. The inset
shows the Nelder-Mead local refinement history corresponding to the solution that
achieves the lowest final $\chi^{2}$ in each case.}}
    \label{top_one_chi2_comparison_for_objective_type}
\end{figure}

In Fig.~\ref{top_one_chi2_comparison_for_objective_type}, we show the evolution of the best (top-1) $\chi^2$ value as a function of the number of valid iterations (\textit{i.e.}, trials that satisfy all imposed constraints) for each optimization type. In both cases we perform 10,000 valid cooling simulations. During the first $\sim10^{3}$ iterations the single-objective run yields larger $\chi^2$ (worse fit) than the multi-objective run. {However, after $10{,}000$ TPE trials, the best $\chi^{2}$ achieved by the single-objective optimization remains lower than that of the multi-objective optimization. To obtain the final best-fit solutions, we further apply a local refinement stage using the Nelder-Mead simplex method, initialized from the top 10 TPE candidates for each optimization. After this refinement, we obtain $\chi^{2}\simeq 30.251$ for the single-objective case and $\chi^{2}\simeq 30.445$ for the multi-objective case.}

{In Fig.~\ref{objective_type_gap_and_tc}(a), we plot the five best gap functions identified
by the TPE optimization for each optimization type, with darker curves corresponding to
lower $\chi^{2}$. Among these, the final best-fit solution obtained after the subsequent
Nelder-Mead refinement is highlighted by a thicker and darker curve.} The black vertical dashed line marks the neutron Fermi momentum at the inner-crust/outer-core boundary, while the green vertical dashed line indicates the neutron Fermi momentum at the center of the star. Thus, values to the right of the vertical green line (shaded region) are not realized inside the neutron star and therefore do not affect the thermal evolution; this explains the comparatively large uncertainty in the $k_2$ parameter observed in Figs.~\ref{parameter_relation_comparison_for_objective_type}(e) and \ref{parameter_relation_comparison_for_objective_type}(j). Consequently, the neutron $^3\mathrm{P}_2$ pairing is physically relevant from the density where its critical temperature falls below that of the neutron ${}^{1}\mathrm{S}_0$ (singlet) pairing up to the stellar center. In other words, for cooling the momentum-dependent critical temperatures---including the ${}^{1}\mathrm{S}_0$ reference---matter more than the bare shape of $\Delta(k_{\mathrm{Fn}})$.

In Fig.~\ref{objective_type_gap_and_tc}(b), we show the corresponding critical temperature for neutron $^3\text{P}_2$ superfluidity as a function of the neutron Fermi momentum. For $k_{\mathrm{Fn}}\gtrsim 1.7~\mathrm{fm}^{-1}$, both optimization types produce nearly flat $T_{\mathrm{c}}$ profiles. The onset of neutron ${}^{3}\mathrm{P}_2$ pairing is then governed by the intersection with the $T_{\mathrm{c}}$ curve for the neutron $^1\text{S}_0$ pairing. We note that, according to Eq.~(\ref{pairing_gap_temp_relation}), the same gap $\Delta$ translates to different critical temperatures $T_{\mathrm{c}}$ for singlet and triplet channels due to the anisotropic reduction factor (triplet $T_{\mathrm{c}}$ is smaller by $\sim 1/\sqrt{8\pi}$). Hence, achieving the same $T_{\mathrm{c}}$ requires a triplet gap roughly five times larger than a singlet gap. With the SFB \cite{Schwenk_2003_SFB} neutron ${}^{1}\mathrm{S}_0$ gap model, we have the critical-temperature profile shown in Fig.~\ref{objective_type_gap_and_tc}(b) represented by a gray dashed curve. {As a result, seemingly different gap shapes in the outer core---\textit{e.g.}, an almost flat model (multi-objective TPE top-2) versus a more bell-shaped model (the Nelder-Mead-refined best-fit curve)---yield very similar $T_{\mathrm{c}}$ intersection
locations and hence comparable $\chi^{2}$ values (\textit{cf.}, $30.437$ vs.\ $30.251$).
} For the present setup ($q\simeq 0.19$, $M_\text{NS}=1.4\,M_{\odot}$), this indicates that the high-density behavior of $T_{\mathrm{c}}$ is more influential for Cas~A NS fitting, while low-density differences in $T_{\mathrm{c}}$ have only minor impact.

It is to mention here that although we adopt the pragmatic rule that the locally dominant neutron pairing channel is set by the larger $T_{\mathrm{c}}$ between ${}^{1}\mathrm{S}_0$ and ${}^{3}\mathrm{P}_2$ pairing, Ginzburg--Landau analyses indicate that ${}^{1}\mathrm{S}_0$ and ${}^{3}\mathrm{P}_2$ condensates can coexist under certain temperature and magnetic-field conditions, potentially smearing the effective onset of triplet PBF and shifting the timing/strength of neutrino emission \cite{Yasui_2020_coexist}. A systematic behavior of such coexistence is an open question and its treatment is beyond the scope of this work. We thus defer investigation of this issue to future extensions of our calibration framework.

We also note that some prior studies (\textit{e.g.}, Ref.~\cite{shternin_2011_nt_SYHHP}) imposed a constant $T_{\mathrm{c}}$ to maximize PBF luminosity. In contrast, our optimization framework favors $T_{\mathrm{c}}$ profiles with a nontrivial momentum dependence, which better reconcile multiple physical quantities affected by ${}^{3}\mathrm{P}_2$ pairing (\textit{e.g.}, heat capacity and thermal conductivity at the core) alongside PBF emissivity. This suggests that a momentum-distributed $T_{\mathrm{c}}$ is more compatible with the Cas~A NS data than a hypothetical uniform $T_{\mathrm{c}}$.

In summary, while the $\chi^{2}$+$\,\mathrm{slope\ diff.}$ multi-objective optimization
samples a broader region of parameter space and provides useful guidance during the
exploratory stage, the final best-fit solution obtained after local refinement achieves a
slightly lower $\chi^{2}$ in the $\chi^{2}$-only single-objective optimization. This outcome
reflects the fact that neutron star cooling curves are intrinsically curved in time, so a
minimum in the slope difference does not necessarily correspond to a minimum in
$\chi^{2}$. Consequently, the subsequent analyses presented below are based on the
$\chi^{2}$-only single-objective formulation, which directly targets the statistic that
aggregates the entire time series.

\begin{figure*}
    \centering
    \includegraphics[width=1.0\textwidth]{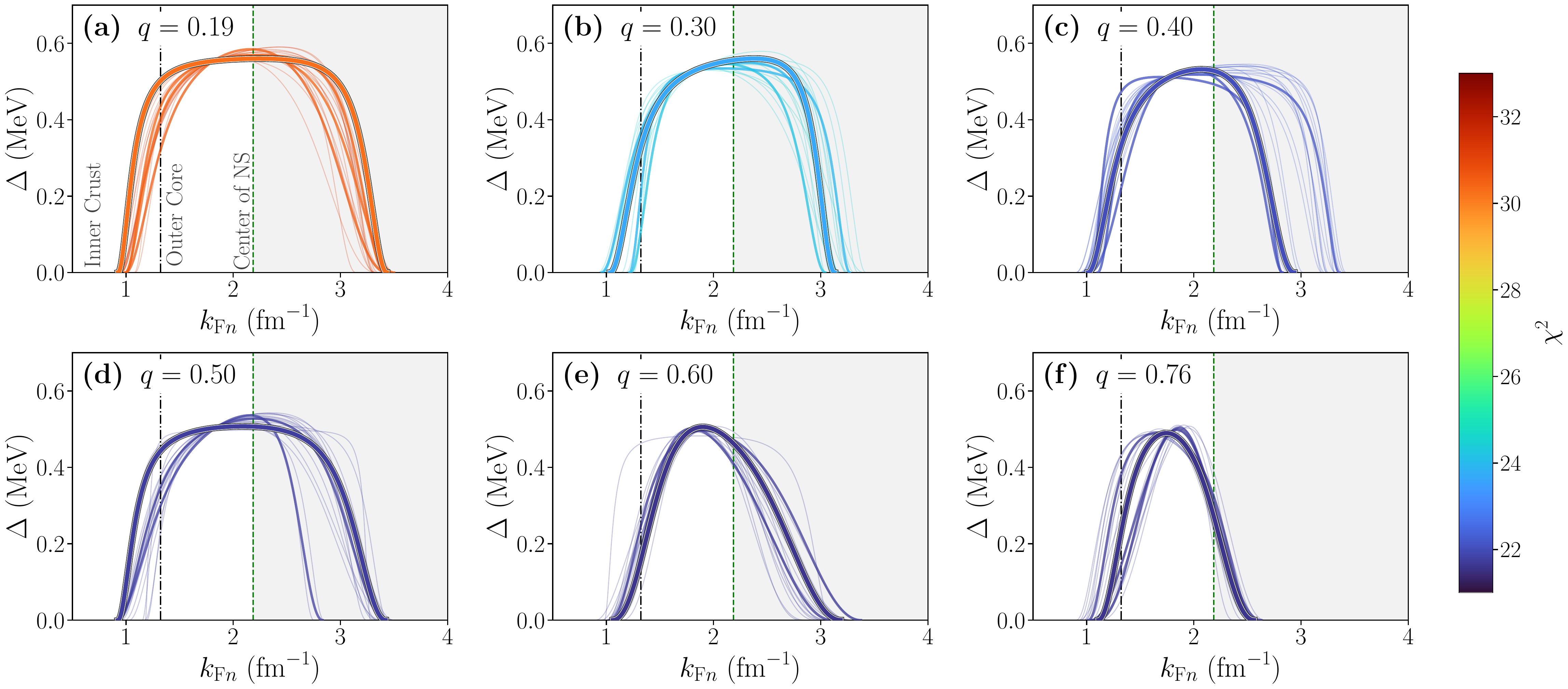}
    \caption{{Top-20 gap functions for each value of $q$ (top-3 highlighted). Curves are colored by the
    resulting cooling curve $\chi^{2}$ against the Cas~A NS data (blue indicates lower
    $\chi^{2}$, red higher). As $q$ increases, the optimized gap shapes exhibit an overall
    tendency toward more localized, bell-shaped profiles in neutron Fermi momentum, although
    a range of widths remains allowed at intermediate $q$.}}
    \label{q_grid_all}
\end{figure*}

\begin{figure}[t]
    \centering
\includegraphics[width=1.0\linewidth]{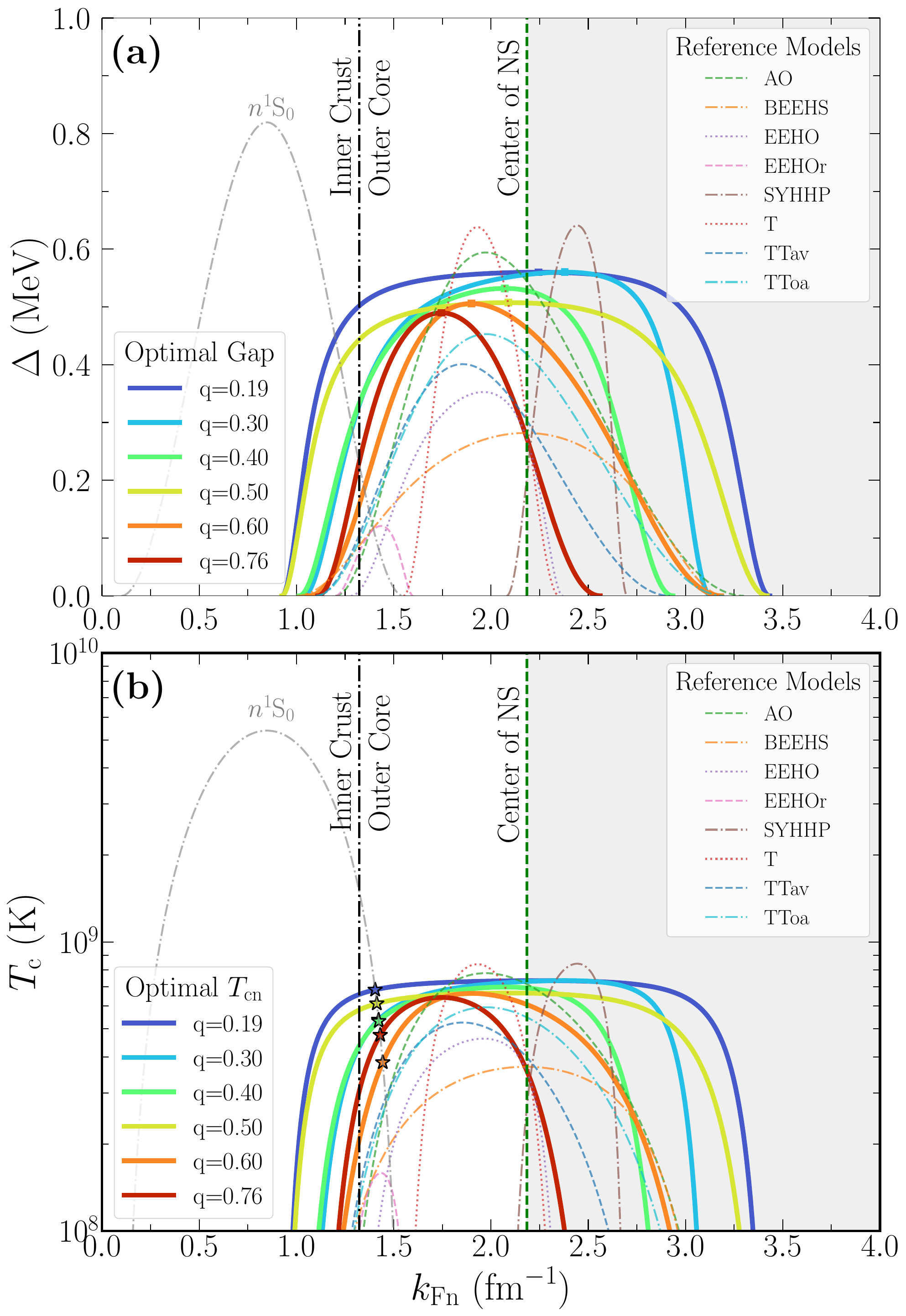}\vspace{-2mm}
    \caption{{ Best-scoring (Nelder-Mead-refined) gap functions (a) and corresponding critical-temperature profiles (b) for each $q$, compared with representative ${}^{3}\mathrm{P}_{2}$ reference models (dotted/dashed) and the neutron ${}^{1}\mathrm{S}_{0}$ SFB model (gray). Apparent crossings with SFB in $\Delta(k_{\mathrm{Fn}})$ at low density are not physically decisive; the relevant intersections occur in $T_{\mathrm{c}}$ [panel (b)]. The optimized parameters are summarized in Table~\ref{tab:best_by_q} (Appendix~\ref{appendix:b}).}}
    \label{q_gap_functions_and_tc_best_one}
\end{figure}


\subsection{$q$ dependence of the parameter optimization at $1.4\,M_\odot$}
\label{subsec:q_at_1p4}

As mentioned in Introduction, there remains uncertainty in the choice of the PBF efficiency factor $q$: while a conventionally-used empirical value is $q\simeq 0.76$ \cite{Page_2009_vector_channel}, a recent microscopic calculation suggests a smaller value of $q\simeq 0.19$ \cite{Leinson_2010}. In this section, treating the PBF efficiency factor $q$ as a free parameter, we explore how the optimized gap properties depend on $q$ while fixing the neutron star mass to the canonical value $M_{\rm NS}=1.4\,M_\odot$. To this end, we perform data-driven
optimizations for $q \in \{0.19,\,0.30,\,0.40,\,0.50,\,0.60,\,0.76\}$. For each value of $q$, we carry out $10{,}000$ valid optimization trials, ensuring a uniform sampling depth across the entire $q$ range and enabling a consistent comparison of the resulting best-fit solutions.

\begin{figure}
    \centering
\includegraphics[width=1.0\linewidth]{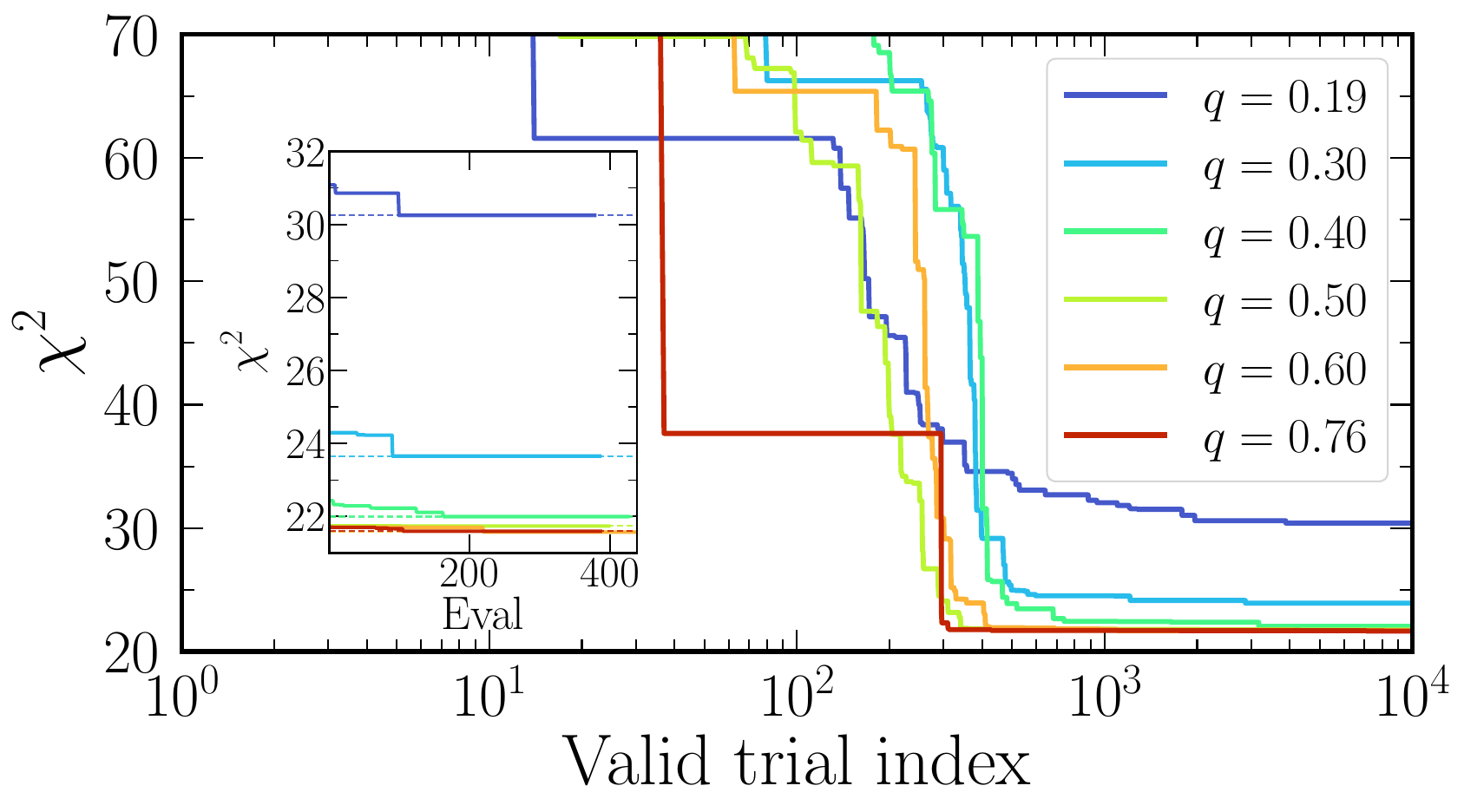}\vspace{-3mm}
    \caption{{Evolution of the best (top-1) $\chi^{2}$ as a function of the valid-trial index for each $q$ (horizontal axis in logarithmic scale). The inset shows the subsequent Nelder-Mead local refinement applied to the best TPE candidate for each $q$, highlighting the final improvement in $\chi^{2}$.}}
    \label{q_chi2_evolution}
\end{figure}

\begin{figure}[b]
    \centering
\includegraphics[width=1.0\linewidth]{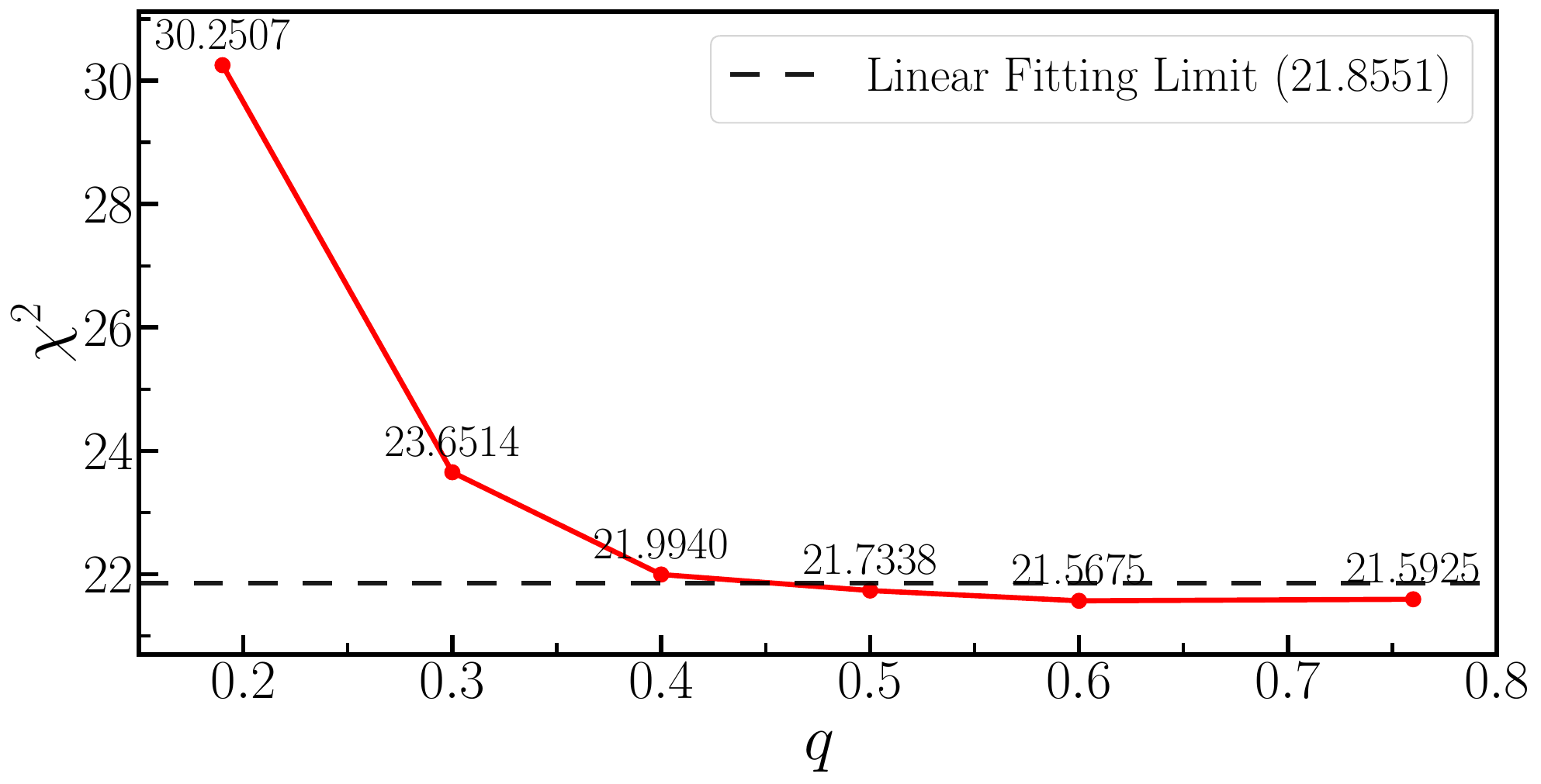}\vspace{-2mm}
    \caption{{Final best (top-1) $\chi^{2}$ as a function of $q$. For $q \ge 0.5$, the best $\chi^{2}$ falls below the horizontal dashed line indicating the $\chi^{2}$ obtained from a linear fit to the Cas~A~NS data, reflecting that theoretical cooling curves are intrinsically curved rather than linear.}}
    \label{q_chi2_relation}
\end{figure}

\begin{figure*}
    \centering
\includegraphics[width=0.8\textwidth]{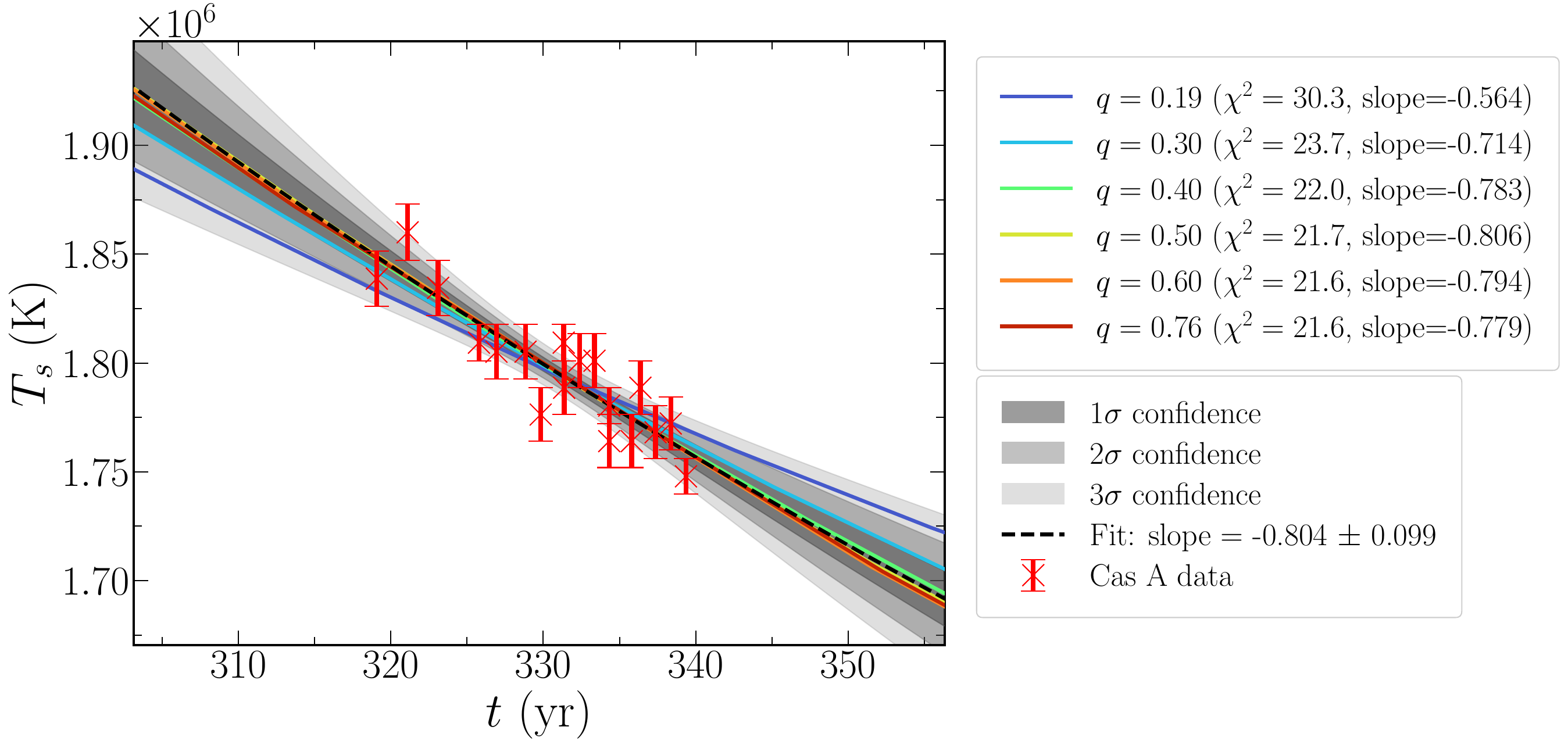}
\caption{{Best (top-1) theoretical cooling curve for each $q$ overlaid on the Cas~A NS surface-temperature measurements (red points). The black dashed line shows the linear fit performed in $\log t$--$\log T_{\mathrm{s}}$ space, and the gray bands indicate the corresponding $1\sigma$, $2\sigma$, and $3\sigma$ confidence intervals. When evaluating $\chi^{2}$, each model curve is allowed to shift in age by up to $\pm19$\,yr.}}
\label{q_cooling_curves}
\end{figure*}

Figures~\ref{q_grid_all}(a)--\ref{q_grid_all}(f) show, in increasing order of $q$, the
top-20 neutron ${}^{3}\mathrm{P}_{2}$ pairing gaps as functions of the neutron Fermi
momentum, highlighting the top-3 results. Each curve is colored according to the
$\chi^{2}$ value of its resulting cooling curve against the Cas~A NS data: lower
$\chi^{2}$ (better score) appears blue, while higher $\chi^{2}$ appears red.
As can be seen from the plots, from Fig.~\ref{q_grid_all}(a) to Fig.~\ref{q_grid_all}(c), the colormap shifts steadily toward blue, indicating improving scores as $q$ increases from $0.19$ to $0.40$. Beyond that, Figs.~\ref{q_grid_all}(d)--\ref{q_grid_all}(f) show no dramatic further color change, suggesting saturation of the best achievable $\chi^2$ near $q \gtrsim 0.5$. In terms of functional form, some of the top models at smaller $q$ exhibit relatively rectangular-like shape at low density, while at larger $q$ the optimized gap shapes tend to be smoother overall. In particular, in Figs.~\ref{q_grid_all}(e)--\ref{q_grid_all}(f) many of the competitive solutions display profiles that are closer to a bell-like form in neutron Fermi momentum, although variations in width persist.

Figures~\ref{q_gap_functions_and_tc_best_one}(a) and \ref{q_gap_functions_and_tc_best_one}(b) show, respectively, {the final best-fit (lowest $\chi^{2}$) gap function obtained after the Nelder-Mead refinement and the
corresponding critical-temperature profile for each $q$.}
Thick solid curves denote our optimized results. We also plot pre-existing pairing gap models listed in Table~\ref{models_and_new_parameters} by thin dotted or dashed curves for comparison. The neutron ${}^{1}\mathrm{S}_0$ gap function (the SFB model) is also shown by a gray dash-dotted line as a reference. Here as well, Fig.~\ref{q_gap_functions_and_tc_best_one}(a) may give the impression (especially at $q\simeq 0.19$) that our optimized gap crosses the SFB gap in the inner crust; however, once mapped to critical temperature---the physically relevant quantity for cooling---the intersections for all $q$ occur in the \emph{outer core}, not the inner crust, as confirmed in Fig.~\ref{q_gap_functions_and_tc_best_one}(b) (intersection points are indicated by star symbols).

{As $q$ increases, Fig.~\ref{q_gap_functions_and_tc_best_one}(b) shows an overall tendency for the onset of neutron ${}^{3}\mathrm{P}_2$ pairing to shift toward higher densities, although non-monotonic behavior can occur at intermediate $q$ owing to the combined effects of the low-density gap structure and the location and height of the gap maximum. In this broader sense, increasing $q$ tends to suppress excessive PBF emissivity at lower densities---where the neutron ${}^{1}\mathrm{S}_0$ (SFB) pairing remains active---thereby avoiding premature cooling. In the high-density region near the stellar center, the optimized triplet $T_{\mathrm{c}}(k_{\mathrm{Fn}})$ profiles for sufficiently large $q$ also tend to develop curvature and attain lower central values, which further mitigates excess PBF emission. Overall, when $q$ is large enough, the resulting optimized gap functions favor more localized, bell-shaped profiles in neutron Fermi momentum and become broadly consistent with representative ${}^{3}\mathrm{P}_2$ models listed in Table~\ref{models_and_new_parameters}. While not strictly monotonic, the optimized solutions also exhibit a tendency for smaller $\Delta_{\max}$ at larger $q$, indicating that the enhanced PBF emissivity is partially offset by a reduction in the overall gap amplitude in the best-fit models.}

{Figure~\ref{q_chi2_evolution} shows the evolution of the best (top-1) $\chi^{2}$ as a function of the valid-trial index (horizontal axis in logarithmic scale) for each value of $q$. The main panel summarizes the optimization histories obtained from the TPE runs, illustrating how the lowest achieved $\chi^{2}$ evolves as sampling proceeds. The inset displays the subsequent Nelder--Mead local refinement performed for the best TPE candidate at each $q$, highlighting the final convergence of the best-fit $\chi^{2}$ values.}

{Figure~\ref{q_chi2_relation} compares the final best (top-1) $\chi^{2}$ values obtained for each value of $q$. The best-fit $\chi^{2}$ decreases as $q$ increases and, for $q \gtrsim 0.5$, falls below the linear-fitting limit $\chi^{2}=21.8551$ derived from the linear fit to the Cas~A NS data. This behavior is expected, since the cooling evolution is intrinsically non-linear in time, and a physically consistent cooling curve can therefore achieve a lower data-wide $\chi^{2}$ than a straight-line approximation. Consistent trends are also visible in the color distributions of the optimized gap functions shown in Figs.~\ref{q_grid_all}(a)--\ref{q_grid_all}(f), where lower $\chi^{2}$ values appear systematically at larger $q$.}

{Finally, in Fig.~\ref{q_cooling_curves}, we compare the best (top-1) theoretical cooling curves for various $q$ values---obtained after Nelder-Mead refinement of the top TPE candidates---with the Cas~A NS observational data.} The red data points are surface temperatures $T_s$ (not the redshifted $T_s^{\infty}$). To obtain the best (top-1) $\chi^2$ at each $q$, we fit with an age offset in $[-19,+19]$~yr to reflect birth-epoch uncertainty; thus, the implied source age differs among $q$ values. For ease of comparison in a single plot, the data are fixed and, instead, the offset is applied to the theoretical cooling curves. The black dashed line shows the linear fit to the Cas~A NS data; the darkest to lightest gray bands indicate the $1\sigma$, $2\sigma$, and $3\sigma$ confidence intervals, respectively. The cooling curves for $q\simeq 0.76$, $q=0.60$, and $q=0.50$ nearly coincide with each other and are indistinguishable, showing slopes almost identical to the linear fit. {For $q=0.40$, the cooling curve is nearly consistent with the linear fit and remains well within the $1\sigma$ band. 
For smaller $q$, the slope becomes progressively shallower: the $q=0.30$ curve lies close to the $1\sigma$–$2\sigma$ boundary (while remaining within $2\sigma$), whereas the $q\simeq 0.19$ case is located near the $2\sigma$–$3\sigma$ boundary (within $3\sigma$).}

For $q \gtrsim 0.4$, the best-fit models reproduce the Cas~A slope within the $1\sigma$ interval, consistent with Ref.~\cite{shternin_2011_nt_SYHHP}. Conversely, even our best $q\simeq 0.19$ solution remains at the $\sim3\sigma$ level and does not enter $2\sigma$, in line with the joint ACIS inference of substantially larger $q$ in Ref.~\cite{Shternin_2022_1.55_0.25}. Taken together, {these trends indicate that the results obtained within our restricted, exploratory setup are not in tension with previous, more model-independent analyses. A more quantitative assessment of the relative roles of PBF microphysics and other physical ingredients, and of their associated uncertainties, requires a systematic Bayesian inference framework in which such elements can be varied and constrained simultaneously.}

\section{Summary and prospect}
\label{Sec:Conclusion}

In this work, we have revisited the rapid cooling of the Cassiopeia~A neutron star (Cas~A NS), focusing on the Cooper-pair breaking and formation (PBF) neutrino emission process, which has long been regarded as one of the most promising explanations. While the PBF process is theoretically well motivated, the actual strength of the process depends on the efficiency factor $q$, whose value remains under debate, and is further complicated by the large model uncertainty of the neutron $^{3}\mathrm{P}_{2}$ pairing gap function. Motivated by these unresolved issues, we have simultaneously accounted for both the uncertainty in $q$ and in the $^{3}\mathrm{P}_{2}$ pairing gap, and conducted data-driven optimizations of them using the observed Cas~A NS data.

One of the key features of this study is the introduction of a novel parametrization of the pairing gap
$\Delta(k_\text{F})$, in which each parameter carries direct physical meaning. Unlike conventional gap
models that rely on phenomenological fits, our parametrization decouples the peak height, location,
width, and asymmetry of the gap, providing both improved interpretability and a natural interface for
machine-learning-type automated applications. As a first exploratory attempt to infer the neutron
$^3\text{P}_2$ pairing gap directly from the Cas~A neutron star observational data, we employ a
data-driven optimization framework based on the tree-structured Parzen estimator (TPE), supplemented
by a local refinement using the Nelder-Mead simplex method. Within this hybrid framework, we perform both
single-objective optimizations based solely on the global $\chi^2$ statistic and multi-objective
optimizations that additionally incorporate the slope difference between the theoretical cooling curve
and the Cas~A~NS data. While the multi-objective formulation explores the parameter space more broadly
and helps guide the search toward physically plausible regions, the single-objective optimization
ultimately achieves slightly lower best-fit $\chi^2$ values. This reflects the intrinsically curved
nature of neutron star cooling trajectories, for which minimizing the slope difference at a
representative epoch does not necessarily guarantee the best global agreement with the entire time
series of the Cas~A NS data.

{Adopting the canonical neutron star mass of $M_{\mathrm{NS}}=1.4\,M_\odot$, we explored the dependence of our results on the PBF efficiency factor $q$ and analyzed the corresponding changes in the optimized neutron $^3\text{P}_2$ pairing-gap functions and critical-temperature profiles. While the evolution of the optimized gap shape with increasing $q$ is not strictly monotonic, a clear overall trend emerges in which larger $q$ values favor smoother, more localized gap profiles that resemble commonly used theoretical models. By comparing the resulting cooling curves with the Cas~A~NS data and with the
slope obtained from linear fits in $\log t$--$\log T_{\mathrm{s}}$ space, we find that models with $q \gtrsim 0.4$ reproduce the observed decline rate within the $1\sigma$ confidence interval. For smaller values of $q$, the cooling slopes become progressively shallower: the $q=0.30$ case approaches the boundary between the $1\sigma$ and $2\sigma$ confidence intervals, while the baseline value $q\simeq 0.19$ resides near the boundary between the $2\sigma$ and $3\sigma$ intervals. This qualitative behavior is consistent with previous studies, although we emphasize that the present conclusion is drawn within a fixed setup that adopts a canonical mass and specific microphysical inputs; a statistically robust and model-independent constraint on $q$ requires a dedicated Bayesian inference that simultaneously accounts for uncertainties in mass, envelope composition, equation of state, and pairing microphysics.}

{Building on this point, we emphasize that the conclusions drawn in this work are necessarily
conditioned on the specific modeling choices adopted here. In particular, we fixed the neutron
${}^{1}\mathrm{S}_{0}$ pairing gap to the SFB model \cite{Schwenk_2003_SFB} and the proton
${}^{1}\mathrm{S}_{0}$ gap to the CCDK model \cite{Elgaroy_1996_ccdk}, while optimizing only the neutron
${}^{3}\mathrm{P}_{2}$ pairing gap. As a result, the present analysis does not exhaust the full space
of microphysical uncertainties, and alternative treatments of singlet pairing gaps may further affect
the inferred cooling behavior. Extending the optimization framework to include the
${}^{1}\mathrm{S}_{0}$ channels therefore constitutes a natural and important next step toward a fully
self-consistent description of nucleon superfluidity and superconductivity in neutron stars.
Likewise, while the present study deliberately excluded direct-Urca processes in order to focus on
PBF-dominated cooling, incorporating such rapid cooling mechanisms within a unified framework remains
an important direction for future work.}

{In addition, the comparison between theoretical cooling curves and the Cas~A~NS data in this study
allowed for a relative age offset within a fixed range ($\pm19$\,yr), which was optimized separately
for each model realization. While this approach is sufficient for the present exploratory analysis,
a Bayesian inference framework would enable a more rigorous treatment by incorporating the neutron
star age as a nuisance parameter directly into the likelihood function. Such an approach would allow
the uncertainty in the Cas~A~NS age to be propagated consistently into the inferred constraints on
cooling microphysics.}

Looking ahead, the physically interpretable parametrization of the pairing gap function introduced
in this study, together with the computational acceleration of neutron star cooling simulations,
provides a natural foundation for systematic Bayesian inference and machine-learning-based analyses.
Such approaches will enable quantitative uncertainty estimates, reveal correlations among
microphysical inputs, and allow simultaneous inference of pairing gaps, stellar structure, and
nuclear equation-of-state parameters. A dedicated follow-up work along this line is in progress \cite{Bayesian_Cooling}. In this way, the methodology developed here opens a pathway
toward next-generation neutron star cooling studies that can more robustly extract information on
dense-matter physics from available observational data.

\section*{Acknowledgments}
We are thankful to Dr.\ Kotaro Uzawa (JAEA) and Dr.\ Yifei Niu (Lanzhou University) for important comments on our presentation on this work that were useful to correctly assess our methodology. We are grateful to Dr.\ Dany P.\ Page (UNAM) and Dr.\ Wynn Ho (Haverford College) for their valuable advice and insightful discussions regarding the implementation and validation of our neutron star cooling code. {We also appreciate the referee of this paper for a number of constructive and insightful suggestions, including the recommendation to incorporate a direct minimization algorithm (the Nelder-Mead method), which significantly improved the robustness and quality of the present analysis.} 
This work is supported by JSPS Grant-in-Aid for Scientific Research, Grants No.~JP23K03410, No.~JP23K25864, {and No.~JP25H01269.}

\appendix

\section{Verifying the parameter distribution of the existing models in the new parametrization}
\label{appendix:a}

In this appendix, we present the parameter distributions of the existing superfluid gap models 
expressed in our new parametrization scheme.

\begin{figure}[t]
    \centering
\includegraphics[width=\linewidth]{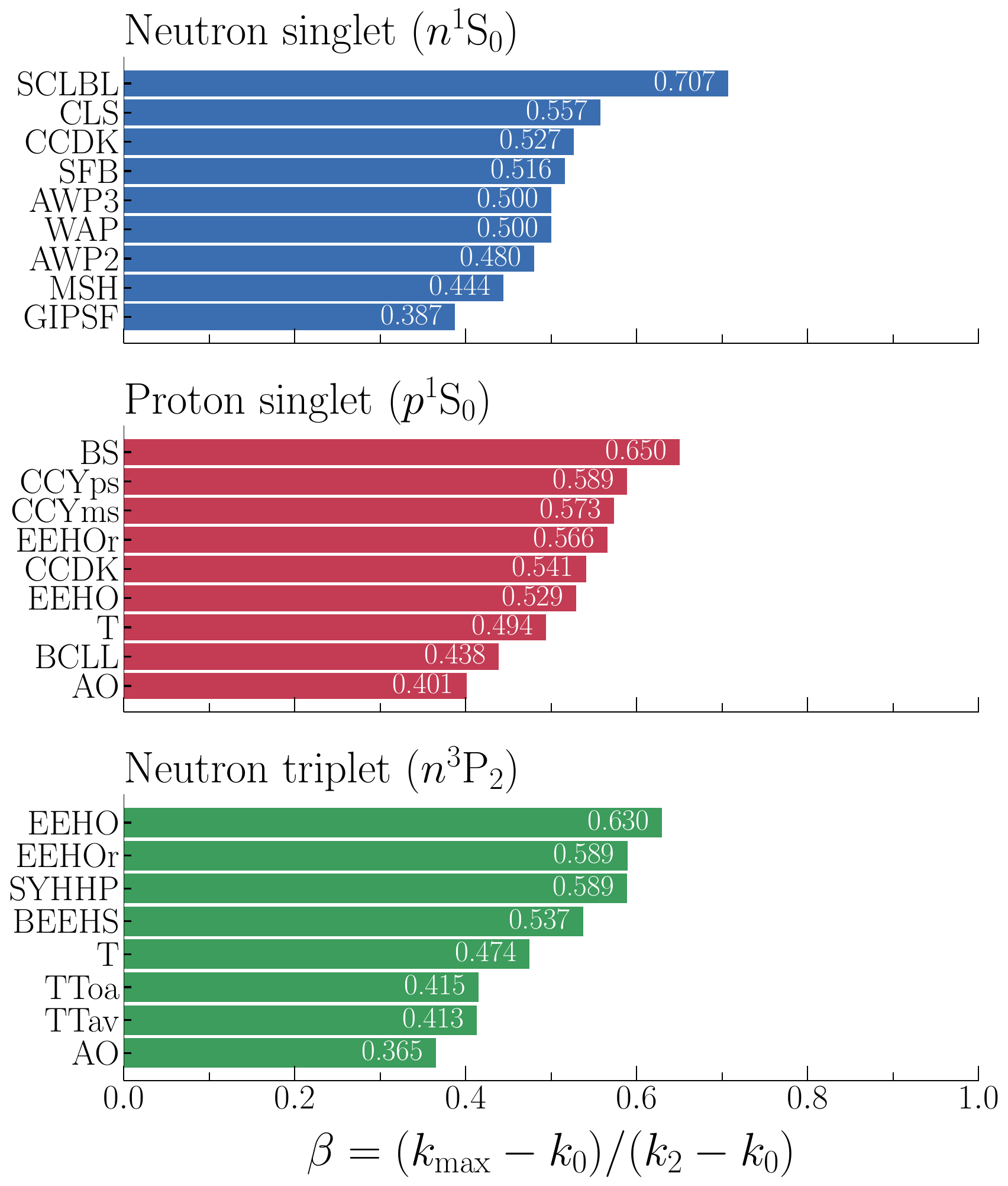}\vspace{-3mm}
   \caption{
Distribution of the parameter $\beta$ for existing superfluid gap models.
Here $\beta = (k_{\max}-k_0)/(k_2-k_0)$ represents the relative position of the Fermi momentum where the gap reaches its maximum within the domain $[k_0, k_2]$.
Blue, red, and green bars correspond to neutron $^1\mathrm{S}_0$, proton $^1\mathrm{S}_0$, and neutron $^3\mathrm{P}_2$ pairings, respectively.
Larger (smaller) $\beta$ values indicate peaks toward the higher (lower) momentum side.
Among the models, $\beta$ ranges from 0.365 (AO) to 0.707 (SCLBL); for neutron $^3\mathrm{P}_2$ gaps, $0.365\! \le\! \beta\! \le\! 0.63$.
In this work, we adopt $0.35\! \le\! \beta\! \le\! 0.65$ to ensure both physical and broad coverage.
}
    \label{appendix_a_beta}
\end{figure}

Figure~\ref{appendix_a_beta} shows the distribution of the parameter $\beta$ for the existing models.
As defined in Section~\ref{sf_gap_model}, $\beta$ represents the relative position of the Fermi momentum 
$k_\mathrm{max}$, where the gap reaches its maximum, within the domain of the gap function $[k_0, k_2]$.
To prevent unphysical gap shapes, we restrict the parameter to the range $0 < \beta < 1$.

Among the existing models, the largest $\beta$ is found for the neutron $^1\mathrm{S}_0$ SCLBL model 
($\beta = 0.707$), indicating that its peak lies toward the right-hand side of the domain,
at about 70\% of the total range.
Conversely, the smallest value occurs in the neutron $^3\mathrm{P}_2$ AO model ($\beta = 0.365$),
implying that its maximum is located toward the lower-momentum side, at roughly 36\% of the range.

Since the present optimization focuses on the neutron $^3\mathrm{P}_2$ gap, 
$\beta$ typically lies within $0.365 \leq \beta \leq 0.63$ for existing $^3\mathrm{P}_2$ models.
To explore a broader yet physically reasonable parameter space, 
we therefore adopt $0.35 \leq \beta \leq 0.65$ in our optimization.

\begin{figure}[t]
    \centering
\includegraphics[width=\linewidth]{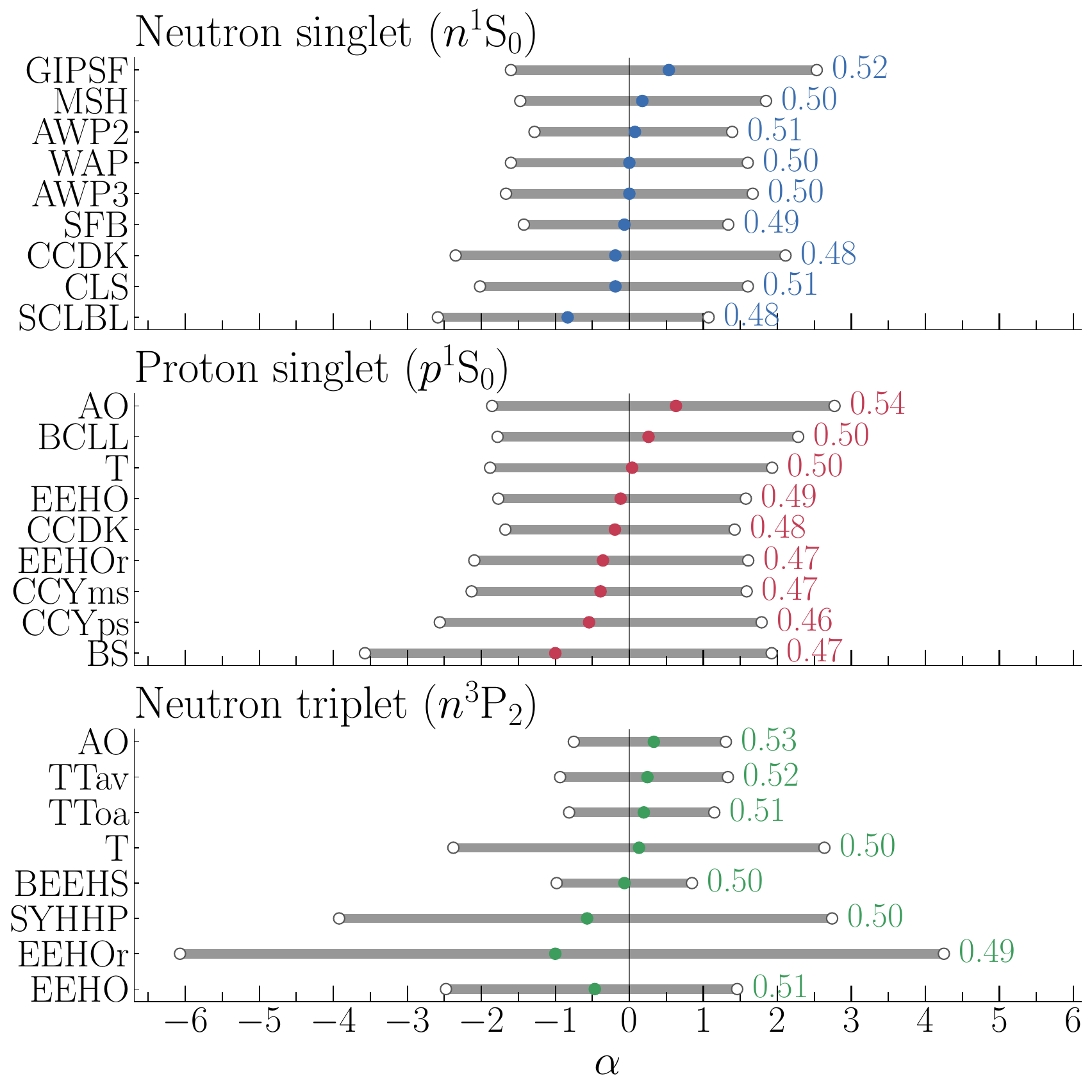}\vspace{-3mm}
\caption{
Distribution of the parameter $\alpha$ for existing superfluid gap models.
The parameter $\alpha$, defined in Section~\ref{sf_gap_model}, controls the asymmetry of the gap and must satisfy Eq.~(\ref{alpha_condition_one}) to avoid divergence within $[k_0,k_2]$.
Gray bars show the allowed ranges, and colored markers denote fitted $\alpha$ values; the numbers indicate their normalized positions within the permitted interval.
Most models yield $\alpha \approx 0.5$, implying nearly symmetric gap shapes:
0.48–0.52 for neutron $^1\mathrm{S}_0$, 0.46–0.54 for proton $^1\mathrm{S}_0$, and 0.49–0.53 for neutron $^3\mathrm{P}_2$.
For our optimization, $\alpha$ is explored within 0.1–0.9 of its allowed range.
}
    \label{appendix_a_alpha}
\end{figure}

Figure~\ref{appendix_a_alpha} shows the corresponding distribution of the parameter $\alpha$.
As defined in Section~\ref{sf_gap_model}, $\alpha$ quantifies the asymmetry of the gap function
and must satisfy Eq.~(\ref{alpha_condition_one}) (or equivalently Eq.~(\ref{al_range}))
to ensure that the function remains finite within $[k_0, k_2]$.

In Fig.~\ref{appendix_a_alpha}, gray bars indicate the allowed ranges of $\alpha$ for each model, 
while colored markers denote the fitted $\alpha$ values obtained from the original model shapes.
The numerical labels to the right of each bar represent the normalized positions of the fitted $\alpha$
within their allowed ranges, defined analogously to $\beta$.
A value of 0.5 corresponds to the midpoint of the allowed range, 
whereas smaller (bigger) values indicate proximity to the lower (upper) bound.

Overall, $\alpha$ values cluster around 0.5 irrespective of the pairing type, 
suggesting that most existing gap functions are nearly symmetric. 
Specifically, $\alpha$ ranges from 0.48–0.52 for neutron $^1\mathrm{S}_0$ models, 
0.46–0.54 for proton $^1\mathrm{S}_0$ models, and 0.49–0.53 for neutron $^3\mathrm{P}_2$ models.
Since the present study aims to explore more general gap shapes, 
we allow $\alpha$ to vary between 0.1 and 0.9 of its physically permitted interval during optimization.

\clearpage

\vspace*{100mm}

\section{{Best-fit gap parameters for different $q$ values}}
\label{appendix:b}

In this appendix, we summarize the best-fit parameters of the new pairing-gap parametrization
obtained from the cooling curve optimization at the canonical neutron star mass $M_{\mathrm{NS}} = 1.4\,M_\odot$.
The parameters listed here correspond to the best-scoring (lowest $\chi^2$) gap functions shown in Fig.~\ref{q_gap_functions_and_tc_best_one}(a).

{Table~\ref{tab:best_by_q} lists the six parameters that fully define the neutron ${}^{3}\mathrm{P}_{2}$ pairing-gap function introduced in Section~\ref{sf_gap_model}: the maximum amplitude $\Delta_{\max}$, the fixed lower and upper endpoints $k_0$ and $k_2$, the location of the peak $k_{\max}$, the width parameter $w$, and the asymmetry parameter $\alpha$. These values are provided as reference data for reproducing the optimized gap functions. Their physical interpretation and $q$-dependent trends are discussed in Section~\ref{subsec:q_at_1p4}.}

\begin{table}[t]
\caption{{Best-fit parameters of the neutron ${}^{3}\mathrm{P}_{2}$ pairing-gap parametrization
for each value of the PBF efficiency factor $q$ at the canonical neutron star mass
$M_{\mathrm{NS}}=1.4\,M_\odot$, obtained after Nelder-Mead refinement.}
}
\vspace{2mm}
\centering
\begin{tabular}{ccccccc}
\hline\hline
\begin{tabular}{c}$q$\end{tabular} &
\begin{tabular}{c}$\Delta_{\max}$\\(MeV)\end{tabular} &
\begin{tabular}{c}$k_0$\\(fm$^{-1}$)\end{tabular} &
\begin{tabular}{c}$k_2$\\(fm$^{-1}$)\end{tabular} &
\begin{tabular}{c}$k_{\max}$\\(fm$^{-1}$)\end{tabular} &
\begin{tabular}{c}$w$\\(fm$^{2}$)\end{tabular} &
\begin{tabular}{c}$\alpha$\\(fm)\end{tabular}
\\ \hline
0.19 & {0.5600} & {0.9257} & {3.4331} & {2.2444} & {6.447} &  {0.4136} \\
0.30 & {0.5602} & {1.0358} & {3.1298} & {2.3792} & {4.069} &  {0.3012} \\
0.40 & {0.5321} & {1.0132} & {2.9358} & {2.0713} & {2.388} &  {0.4899} \\
0.50 & {0.5074} & {0.9245} & {3.4209} & {2.0892} & {3.510} &  {0.5174} \\
0.60 & {0.5059} & {1.0603} & {3.1858} & {1.8994} & {0.668} & {-0.1104} \\
0.76 & {0.4898} & {1.1086} & {2.5614} & {1.7447} & {1.801} &  {0.4964} \\
\hline\hline
\end{tabular}
\label{tab:best_by_q}
\end{table}

\clearpage

\bibliography{ref}

\end{document}